\documentclass[11pt, aps, superscriptaddress,amsmath,amssymb]{revtex4-1}
\usepackage{graphicx} % Include figure files
\usepackage{dcolumn}  % Align table columns on decimal point
\usepackage{bm}       % bold math
\usepackage{amsfonts}
\usepackage{color}

\usepackage{soul,xcolor}
\usepackage{ulem}
\usepackage{color}

\begin{document}

\preprint{}

\title{High-Harmonic Spectroscopy of Coherent Lattice Dynamics in Graphene}

\author{Navdeep Rana}
\affiliation{%
Department of Physics, Indian Institute of Technology Bombay,
            Powai, Mumbai 400076  India}

\author{M. S. Mrudul}
\affiliation{%
Department of Physics, Indian Institute of Technology Bombay,
            Powai, Mumbai 400076  India}

\author{Daniil Kartashov}
\affiliation{%
Institute of Optics and Quantum Electronics, Friedrich-Schiller University Jena, Max-Wien-Platz 1, 07743 Jena, Germany}
\author{Misha Ivanov}     
\affiliation{%
Max-Born Stra{\ss}e 2A, D-12489 Berlin, Germany }

\author{Gopal Dixit}
\email{gdixit@phy.iitb.ac.in}
\affiliation{%
Department of Physics, Indian Institute of Technology Bombay,
            Powai, Mumbai 400076  India}
\date{\today}

%\pacs{}

%%%%%%%%%%%%%%%%% END OF PREAMBLE %%%%%%%%%%%%%%%%

\begin{abstract}
High-harmonic spectroscopy 
of solids is a powerful tool, which provides access to both electronic structure and ultrafast electronic response of solids, from their band structure  
and density of states, to  phase transitions, 
including the emergence of the topological edge states, 
to the PetaHertz electronic response.  
However, in spite of these successes,  high harmonic spectroscopy has hardly been applied  to analyse the role of coherent femtosecond lattice vibrations in the attosecond electronic response. Here we  study coherent phonon excitations in monolayer graphene to show how high-harmonic spectroscopy can be used to detect the influence of coherent lattice dynamics, particularly longitudinal and transverse optical  
phonon modes, on the electronic response.  
Coherent excitation of the in-plane  phonon modes results 
in the appearance of sidebands in the spectrum  
of the emitted harmonic radiation. We show that the spectral positions and the polarisation 
of the sideband emission offer a sensitive probe of the 
dynamical symmetries associated with the excited 
phonon modes. 
Our work brings the key advantage of high harmonic spectroscopy -- the combination of sub-femtosecond to tens of femtoseconds temporal resolution -- to the problem of  probing phonon-driven electronic response and its dependence on the dynamical symmetries in solids. 
\end{abstract}

\maketitle

\section{Introduction} 
Strong-field driven high-harmonic generation (HHG) is a  nonlinear frequency up-conversion process, which emits radiation at integer multiples of the incident laser frequency~\cite{ferray1988multiple}.  
Taking advantage of major technical advances in mid-infrared sources, the pioneering experiments \cite{ghimire2011observation} have extended HHG from gases to solids, stimulating intense research into probing electron dynamics in solids on the natural timescale. Today, high-harmonic spectroscopy has been employed to probe different static and dynamic  properties of solids, such as  band dispersion~\cite{vampa2015all, luu2015extreme, lanin2017mapping, pattanayak2019direct}, density of states~\cite{tancogne2017impact},  band defects~\cite{mrudul2020high, pattanayak2020influence}, valley pseudospin~\cite{mrudul2021light, jimenez2020lightwave, langer2018lightwave, mrudul2021controlling}, 
Bloch oscillations~\cite{schubert2014sub},
topology and light-driven phase transitions, 
including strongly correlated systems ~\cite{bauer2018high, silva2018high, bai2020high, imai2020high, borsch2020super, baykusheva2021all, bharti2022high, pattanayak2022role, chacon2020circular, shao2022high}, and even combine attosecond temporal with pico-meter spatial resolution of electron trajectories in lattices~\cite{lakhotia2020laser}. 

Availability of mid-infrared light sources also enables coherent excitation of a desired  
phonon mode by tuning the polarisation and frequency of the laser pulse  ~\cite{mankowsky2016non}.  Yet, the analysis of the effect of coherent lattice dynamics  
on high harmonic generation in solids appears 
lacking, apart from a lone
experiment~\cite{hollinger2019high}. 
This situation stands in stark contrast to molecular gases, 
where  high-harmonic spectroscopy  has been 
extensively employed to probe nuclear motion 
in various molecules~\cite{patchkovskii2009nuclear, wagner2006monitoring, le2012theory, baker2006probing, lein2005attosecond, worner2011conical}. 
Present work aims to fill this gap and highlight some of the capabilities offered by high harmonic spectroscopy in time-resolving 
the interplay of femtosecond lattice and attosecond electronic motions. 
Such interplay is essential for many fundamental phenomena, including  thermal conductivity~\cite{Niedziela_2019}, optical reflectivity~\cite{Dove1993, katsuki2013all}, structural phase transition~\cite{bansal2020magnetically, hase2015femtosecond}, heat capacity~\cite{Bansal_2018}, and optical properties~\cite{Fultz2010, gambetta2006real}. 

Various spectroscopic methods have been developed to excite and probe phonons, see e.g. ~\cite{dhar1994time, debnath2021coherent, graf2007spatially, virga2019coherent, koivistoinen2017time, rana2021four, brown2019direct, flannigan2018electrons, gierz2015phonon, moulet2017soft, geneaux2020attosecond}, but their temporal resolution is limited by the length of the pulses used. 
Large coherent bandwidth of high harmonic signals offers sub-laser-cycle temporal resolution and 
the possibility to time-resolve the impact of lattice distortions on the faster electronic response.

One difficulty in tracking lattice vibrations via highly nonlinear optical response stems from their 
small amplitude. If the corresponding changes in both the band structure and  couplings are similarly small, the high-harmonic response hardly changes. Yet, large distortions are not needed if the excited phonon mode 
dynamically changes the symmetry of the unit cell.
Here we show how coherent phonon dynamics and the associated changes in the lattice symmetry are encoded  
in the electronic response and the harmonic signal, and how the sub-cycle temporal resolution inherent in the harmonic signal can be used to track the interplay of electronic and lattice dynamics. 

\begin{figure}
	\includegraphics[width= 0.8 \linewidth]{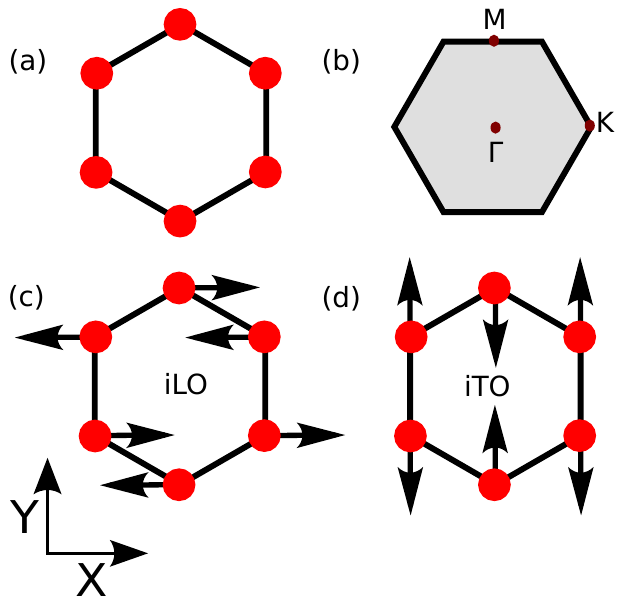}
	\caption{Hexagonal honeycomb structure of graphene and associated in-plane phonon modes. 
	(a) Real-space structure of graphene. (b)  Brillouin zone in momentum-space with 
	$\mathsf{\Gamma, M,}$ and $\mathsf{K}$  as the high symmetry points. 
	(c) and (d) are the sketches of atomic vibrations associated with the degenerate E$_{2g}$ phonon 
	modes in real-space. Here, modes are labelled as (c) in-plane longitudinal optical ($\textsf{iLO}$) 
	phonon mode and (d) in-plane transverse optical ($\textsf{iTO}$) phonon mode, respectively.}\label{graphene}
\end{figure}

We analyse monolayer graphene, which 
belongs to $\textbf{D}_{6h}$ point group symmetry, see Fig.~\ref{graphene}(a).  It exhibits six phonon branches: three optical and three acoustic. Here we focus on the former.
Out of the three optical phonon modes, one  is out-of-plane,  the two others are in-plane modes.     
We will consider only the in-plane modes. 
The lattice vibrations corresponding to the in-plane Longitudinal Optical ($\textsf{iLO}$), and the in-plane Transverse Optical ($\textsf{iTO}$) E$_{2g}$ modes are shown in Fig.~\ref{graphene}(c) and \ref{graphene}(d), respectively.
The two modes are  degenerate at the $\mathsf{\Gamma}$-point, with the phonon frequency equal to 194 meV (oscillation period $\sim$ 21 femtoseconds~\cite{kim2013coherent}); both are Raman active and can be excited with a resonant pulse pair or
impulsively by a short pulse with bandwidth covering 
194 meV.
Moreover, it is possible to selectively excite either  $\textsf{iLO}$ or $\textsf{iTO}$ coherent phonon mode by tuning the polarisation of the pump pulse either along   $\mathsf{\Gamma-K}$ or $\mathsf{\Gamma-M}$ direction, respectively.
 
Coherent lattice dynamics should in general introduce 
periodic modulations of the system parameters and thus of its high-harmonic response. In the frequency domain, such modulations 
add sidebands to the main peaks in the harmonic spectrum. 
We shall see that their position and polarisation encode the information about the frequency and the symmetry of the excited phonon mode, respectively.   

\section{Theoretical Method}
Carbon atoms are arranged at the corners of a hexagon in the honeycomb lattice of the graphene.  
The unit-cell of graphene has a two-atom basis, usually denoted as A and B atoms.
The corresponding Brillouin zone in momentum space is shown in Fig.~\ref{graphene}(b), 
where $\mathsf{ \Gamma}$, $\mathsf{M}$, and $\mathsf{K}$ are the high-symmetry points. 
In our convention, the zigzag and  armchair directions of graphene
are along  $\mathsf{X}$-axis ($\mathsf{\Gamma}-\mathsf{K}$ direction) and 
$\mathsf{Y}$-axis ($\mathsf{\Gamma}-\mathsf{M}$ direction), respectively. 

The electronic ground-state of the graphene is  described by the nearest-neighbour tight-binding approximation and  the corresponding Hamiltonian is written as
\begin{equation}
\mathcal{\hat{H}}_\textbf{k} = - \gamma_{0}\sum_{i\in nn} ~e^{i\textbf{k}\cdot \textbf{d}_i} \hat{a}_\textbf{k}^{\dagger} \hat{b}_\textbf{k} + \textrm{H. c.} \label{eq:tb}
\end{equation}
Here, the summation is over the nearest neighbour atoms. 
 $\gamma_0$ is the nearest-neighbour hopping energy, which is chosen to be 2.7 eV.
 $\textbf{d}_i$ is the separation vector between an atom with its nearest neighbour, such that $|\textbf{d}_i| = a$ = 1.42 \AA~is the inter-atomic distance, for a lattice parameter $a_0$ of 2.46 \AA. 
 $\hat{a}_k^{\dagger} (\hat{b}_k)$  is creation (annihilation) operator for  atom A (B) in the unit cell. 
 The low-energy band-structure of graphene is obtained by solving Eq.~(\ref{eq:tb}) and 
 has zero band-gap and exhibits linear dispersion at $\mathsf{K}$-points in the Brillouin zone.

We treat lattice dynamics classically, and assume that atoms perform harmonic oscillations for short displacements from their equilibrium positions. The displacement vector for a particular phonon mode is expressed as 
\begin{equation}
	\textbf{q}(t)=   \textrm{q}_{0}~\hat{\textbf{e}}~\rm{Re}\left( e^{i\omega_{ph}t}\right).
\end{equation}
Here, $\textrm{q}_{0}$  is the maximum displacement of an atom from its equilibrium position,  
$\omega_{\textrm{ph}} = $ 194 meV is the energy of the E$_{2g}$ phonon mode ,  
and $\hat{\textbf{e}}$ is the normalised eigenvector for a particular phonon mode. 
From Figs.~\ref{graphene}(c) and \ref{graphene}(d), it is clear that $\hat{\textbf{e}}_{\textsf{iLO}}$ = $[1,0,-1,0]/\sqrt{2}$ and $\hat{\textbf{e}}_{\textsf{iTO}}$ = $[0,1,0,-1]/\sqrt{2}$, in which the first (last) two elements are components of A (B) atom. 

Due to coherent phonon excitations, lattice dynamics causes temporal variations in the relative distance 
between atoms ($\textbf{d}_i$). In this case, the corresponding time-dependent Hamiltonian within the tight-binding approximation can be written as~\cite{mohanty2019lazy, rodriguez2021direct, wang2014topological}
\begin{equation}
\mathcal{\hat{H}}_\textbf{k}(t) =  - \gamma(t)\sum_{i\in nn} e^{i\textbf{k}\cdot \textbf{d}_i(t)} \hat{a}_\textbf{k}^{\dagger} \hat{b}_\textbf{k} + \textrm{H. c.}
\label{eq:tth}
\end{equation}
Here, the hopping energy is modelled as an exponentially decaying function of the 
relative displacement between nearest-neighbour atoms as 
$\gamma(t)$ = $\gamma_0~e^{-(|\textbf{d}_{i}(t)|-a)/\delta}$, in which $\delta$ is the 
width of the decay function chosen to be 0.184$a_0$~\cite{moon2013optical}.

The interaction among  laser, electrons and coherently excited phonon mode in graphene is modelled by solving following equations of the single-particle density matrix. 
By updating the modified Hamiltonian as a result of the lattice dynamics,  
semiconductor Bloch equations in co-moving frame $|n,\textbf{k}+\textbf{A}(t)\rangle$, is extended and 
equations of motion read as
\begin{subequations}
\begin{align}
    \frac{d}{dt}\rho_{vv}^{\textbf{k}} &= i\textbf{E}(t)\cdot \textbf{d}_{vc}(\textbf{k}_t, t)\rho_{cv}^{\textbf{k}} + \textrm{c.c.} \\
    \frac{d}{dt}\rho_{cv}^{\textbf{k}} &= \left[-i\varepsilon_{cv}(\textbf{k}_t, t)-\frac{1}{T_2}\right]\rho_{cv}^{\textbf{k}} + i\textbf{E}(t)\cdot\textbf{d}_{cv}(\textbf{k}_t, t)\left[\rho_{vv}^{\textbf{k}}-\rho_{cc}^{\textbf{k}}\right].
\end{align}
\label{SBE}
\end{subequations} 
Here, $\textbf{E}(t)$ and $\textbf{A}(t)$ are, respectively, the electric field and the vector potential corresponding to the laser field, which are related as  $\textbf{E}(t)$ = $-d\textbf{A}(t)/dt$, and
$\textbf{k}_t$ is the shorthand notation for $\textbf{k} + \textbf{A}(t)$. 
$\varepsilon_{cv}(\textbf{k})$ and $\textbf{d}_{cv}(\textbf{k})$ are, respectively, the band-gap energy and dipole matrix elements between valence and conduction bands at $\textbf{k}$. $\textbf{d}_{cv}(\textbf{k})$ is defined as $\textbf{d}_{cv}(\textbf{k}) = i\langle c,\textbf{k} |\nabla_\textbf{k}|v,\textbf{k}\rangle$. Also, $\rho_{cc}^{\textbf{k}}(t) = 1 - \rho_{vv}^{\textbf{k}}(t)$, and $\rho^{\textbf{k}}_{vc}(t) = \rho^{\textbf{k}*}_{cv}(t)$.  

A phenomenological term to take care of the interband decoherence is added with a constant dephasing time $T_2$. 
We calculate the matrix elements at each time-step during temporal  
evolution of the coherently excited phonon mode, which results in the additional time-dependence in the matrix elements. 
As long as the maximum displacement of the atoms are small and the time-step is too small compared to the phonon time-period, the matrix elements at consecutive time-steps are smoothly updated.  
	
We solve the coupled differential equations described in Eq.~(\ref{SBE}) 
using the fourth-order Runge-Kutta method with a time-step of 0.01 fs. 
 We sampled the Brillouin zone 
with 251$\times$251 grid.  
The current at any \textbf{k} point in the Brillouin zone is defined as
\begin{equation}
\textbf{J}(\textbf{k}, t)   = \sum_{m,n  \in \{c,v\} } \rho_{mn} ^{\textbf{k}} (t)  \textbf{p}_{nm}(\textbf{k}_t, t).
\end{equation}
Here, $\textbf{p}_{nm}$ are the momentum matrix-elements defined as $\textbf{p}_{nm}(\textbf{k}) = \langle n,\textbf{k}|\nabla_\textbf{\textbf{k}}\mathcal{\hat{H}}_\textbf{k}| m,\textbf{k}\rangle$. 
The total current, \textbf{J}(t) can be calculated by integrating $\textbf{J}(\textbf{k}, t)$ over 
the entire Brillouin zone. 

The high-harmonic spectrum is simulated as
\begin{equation}
\mathcal{I}(\omega) = \left| \mathcal{FT} \left(  \dfrac{\rm{d}}{{\rm d}t} \textbf{J}(t)    \right) \right|^2.
\end{equation}
Here, $\mathcal{FT}$ stands for the Fourier transform. 

High-order harmonics are generated from monolayer graphene, with or without coherent lattice dynamics, using a linearly polarised pulse with a  wavelength of 2.0 $\mu$m and peak intensity of 
1$ \times 10^{11} $ W/cm$^2$. 
The pulse is 100 fs long and has a sin-squared envelope.  The laser parameters used in this work 
are below the damage threshold of graphene~\cite{currie2011quantifying}. 
Similar laser parameters have been used to investigate  electron dynamics in graphene via intense laser pulse
~\cite{heide2018coherent, higuchi2017light, yoshikawa2017high}. 
The value of the 
dephasing time $T_2 = $  10 fs is used throughout in this work~\cite{mrudul2021high}. 
The observations we made here are consistent for other values of $T_2$ in the range 5 - 30 fs. 
Both in-plane E$_{2g}$ phonon modes are considered here. 
Results presented in this work correspond to a maximum 0.03a$_0$ 
displacement of  atoms from their 
equilibrium positions during coherent lattice dynamics. However, our findings remain valid for  displacements ranging from 0.01a$_0$ to  0.05a$_0$ with respect to the equilibrium positions.
\section{Results and Discussion}

%\subsection{Sensitivity of High-Harmonic Generation to Coherent Lattice Dynamics}

\begin{figure}
\includegraphics[width=\linewidth]{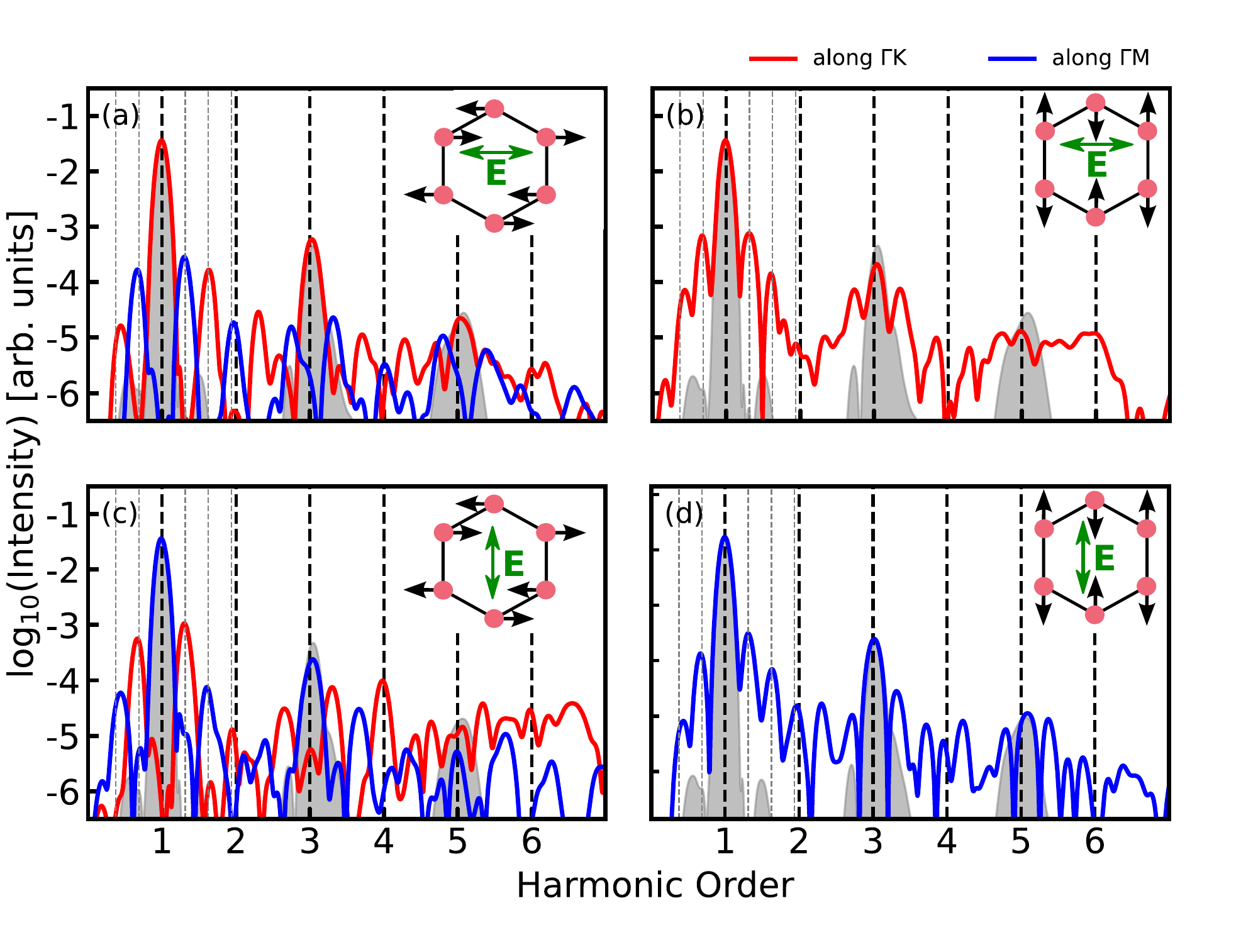}
\caption{High-harmonic spectra of monolayer graphene with and without coherent lattice dynamics.
(a) and (c) High-harmonic spectra corresponding to the coherent $\textsf{iLO}$ E$_{2g}$ phonon mode 
and the probe harmonic pulse is polarised along $\mathsf{\Gamma K}$ and $\mathsf{\Gamma M}$ directions, respectively. 
(b) and (d) Same as (a) and (c) except $\textsf{iTO}$ E$_{2g}$ phonon mode is coherently excited. 
In all the cases, sidebands corresponding to the first harmonic are marked at frequencies 
($\omega_{0} \pm  m \omega_{\textrm{ph}}$), where $\omega_0$  is the frequency of the harmonic generating 
probe pulse, and  $\omega_{\textrm{ph}}$ is phonon frequency. 
The harmonics with grey shaded area are the reference spectra and 
represent the spectra of graphene without phonon excitation. 
The unit cell of the graphene with the corresponding phonon eigenvector and polarisation of the harmonic generating probe pulse are shown in the respective insets.  
Red (blue) colour corresponds the polarisation of emitted radiation parallel (perpendicular) 
to the polarisation of harmonic generating 
probe pulse.}
\label{HHGlattice}
\end{figure}

High-harmonic spectra for monolayer graphene, with and without coherent lattice dynamics, are 
presented in Fig.~\ref{HHGlattice}. The spectrum corresponding to the graphene, without lattice dynamics, 
is shown by grey shaded area as a reference.  
Owing to the inversion symmetry of the graphene, the reference spectrum in grey colour 
exhibits only odd harmonics (consistent with earlier reports, e.g. Refs.~\cite{mrudul2021high,yoshikawa2017high,al2014high}.) 

We assume that coherent phonon dynamics is excited prior to a high harmonic probe.
When one of the E$_{2g}$ phonon modes in graphene is coherently excited, 
the harmonic spectra display sidebands along with the main odd harmonic peaks as reflected from Fig.~\ref{HHGlattice}. The energy difference between the adjacent sidebands matches the phonon energy ($\omega_{\textrm{ph}}$). 
The sideband intensity is sensitive to the phonon amplitude but is clearly visible already for amplitudes above 0.01 of the lattice constant. 
Here we present the case of the amplitude equal to 0.03 of the lattice constant.

As E$_{2g}$ phonon modes preserve inversion centre, 
only odd harmonics are generated. 
When the coherent $\textsf{iLO}$ mode and the probe harmonic pulse (along  
$\mathsf{\Gamma-K}$) are in the same direction,  
the even-oder sidebands are polarised along $\mathsf{\Gamma-K}$ (red colour), whereas the 
odd-order sidebands are polarised perpendicular to $\mathsf{\Gamma-K}$ (blue colour), i.e., 
along $\mathsf{\Gamma - M}$ direction [see Fig.~\ref{HHGlattice}(a)]. 
When the polarisation of the probe pulse changes from
$\mathsf{\Gamma-K}$ to $\mathsf{\Gamma - M}$ direction, 
the polarisation of the sidebands  remains the same with respect to the laser polarisation. 
In this case, 
the even-oder sidebands are polarised along $\mathsf{\Gamma-M}$  (blue colour), whereas
odd-order sidebands are polarised along $\mathsf{\Gamma-K}$ (red colour) [see Fig.~\ref{HHGlattice}(c)].
In both the cases, the main harmonic peaks are always polarised along the direction of the probe pulse. 

The situation is simpler in the case of coherent $\textsf{iTO}$ mode excitation. 
Both the main harmonic peaks and the 
sidebands are polarised along the direction of the probe pulse 
[see Figs.~\ref{HHGlattice}(b) and (d)]. 
Thus, we see that the polarisation of the 
sidebands yields information about the
symmetries of the excited phonon modes.

\begin{figure}
	\includegraphics[width=\linewidth]{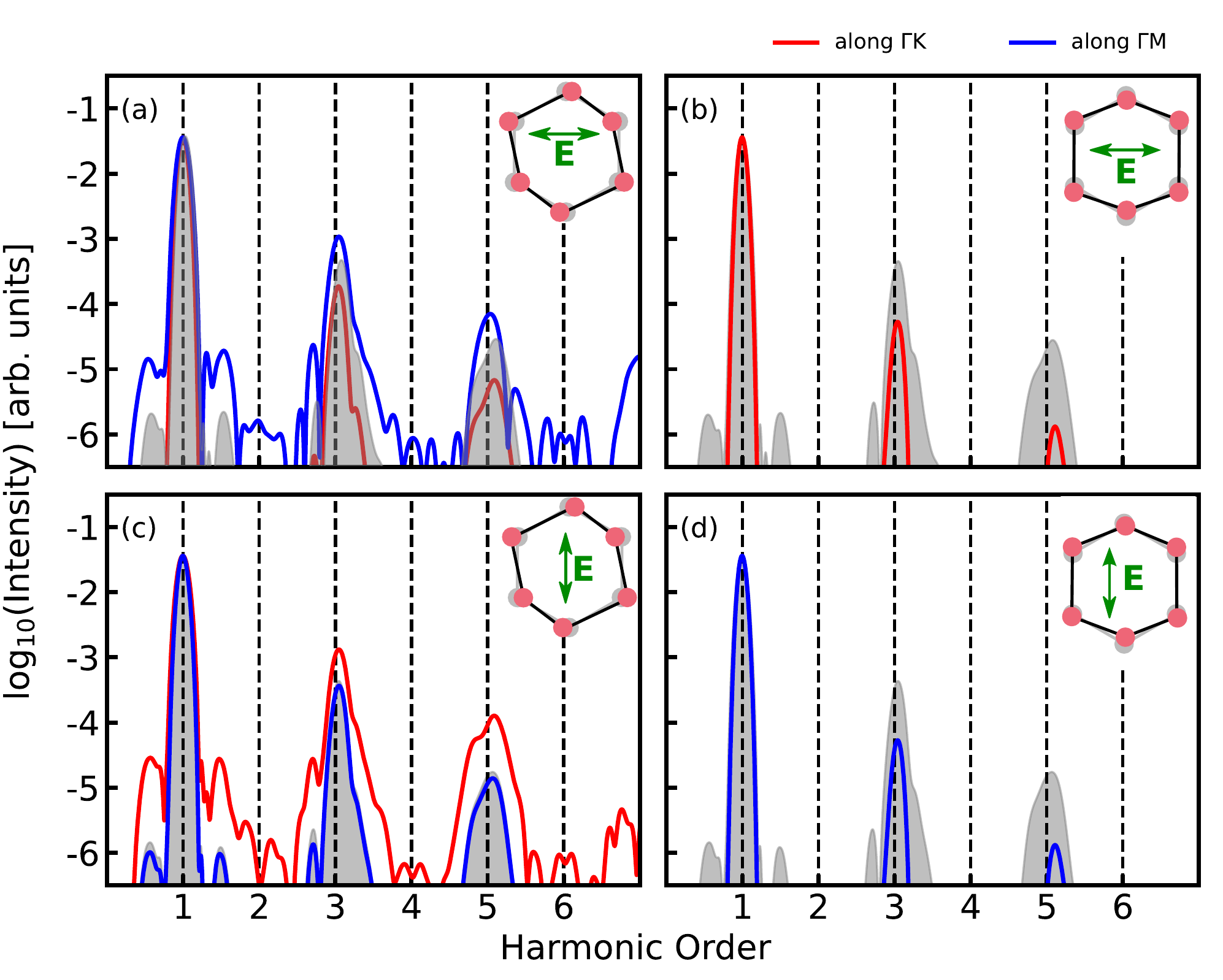}
	\caption{High-harmonic spectra of  monolayer deformed graphene. 
	(a) and (c) When the atoms in graphene are maximally displaced, from their equilibrium position,  
	along  $\textsf{iLO}$  phonon mode. (b) and (d) Similar to (a) and (b) but atoms are displaced 
	along $\textsf{iTO}$   phonon mode. The harmonic spectrum of undeformed graphene is shown 
	in grey shaded area for reference. The unit cell of the deformed graphene lattice and the 
	polarisation of the harmonic generating 
probe pulse are shown in the respective insets. Red (blue) colour corresponds the polarisation of emitted radiation parallel (perpendicular) 
to the polarisation of harmonic generating 
probe pulse.}
	\label{HHGdeformed}
\end{figure}

We now investigate 
how the dynamical changes in symmetries  
differ from similar static variation in the  
high-harmonic spectra. 
Consider the static case  with the   
maximum displacement of atoms, along a
particular phonon mode direction, 
3$\%$ of the lattice parameter from their equilibrium positions. 
Figure~\ref{HHGdeformed} compares high-harmonic spectra for the statically-deformed and undeformed graphene
(grey color). 
The probe polarisation is  along  $\mathsf{\Gamma-K}$ and $\mathsf{\Gamma-M}$ directions in the top and bottom panels of Fig.~\ref{HHGdeformed}, respectively.  
 
When the graphene is deformed along the $\textsf{iLO}$ phonon mode, 
odd harmonics are generated along parallel and perpendicular directions with respect to the laser polarisation
as shown in Figs.~\ref{HHGdeformed}(a) and \ref{HHGdeformed}(c), respectively.  
However, only odd harmonics,  parallel to the laser polarisation, are  generated  when graphene  is deformed  in accordance with $\textsf{iTO}$-mode [see Figs.~\ref{HHGdeformed}(b) and (d)]. 

The emergence of parallel and the perpendicular components in the first case and 
the parallel component in second case 
can be explained as follows: 
the monolayer graphene has $\sigma_{x}$ and $\sigma_{y}$ symmetry planes, 
in addition to the inversion centre. 
When the polarisation of the probe laser is along the high symmetry direction ($\mathsf{\Gamma-K}$ or $\mathsf{\Gamma-M}$), there is no perpendicular component of the current.
However, if the polarisation of the probe pulse 
is along any other than these high-symmetry directions, 
symmetry constraints allow the generation 
of  odd harmonics perpendicular to the direction of the laser polarisation. Recently, same symmetry concept is employed  
in twisted bilayer graphene to correlate the twist angle with its high-harmonic spectrum~\cite{du2021high}. 
It is straightforward  to see that the distortion due to the $\textsf{iLO}$ phonon mode breaks the symmetries of the reflection planes in monolayer graphene. 
The absence of the reflection symmetry planes along \textsf{X} and \textsf{Y} directions 
guarantees the generation of harmonics in both   $\mathsf{\Gamma-K}$ and $\mathsf{\Gamma-M}$ directions  as shown in Figs.~\ref{HHGdeformed}(a) and (c). 
On the other hand, $\textsf{iTO}$ phonon mode preserves both the symmetry planes and as a results harmonics along the laser polarisation are only allowed [Figs.~\ref{HHGdeformed}(b) and (d)].

In short, the presence or absence of the perpendicular current is a result of the transient breaking of 
the symmetry planes, which can be correlated to the results in Fig.~\ref{HHGlattice}.  To understand the mechanism behind the sideband generations and associated polarisation properties during a coherent lattice dynamics, we need to consider the changes in the symmetries dynamically  during the probe pulse. 

To understand the symmetry constraint on the polarisation of sidebands, 
let us consider dynamical symmetries (DSs) of the system, accounting for the coherent lattice dynamics and the probe pulse.
We  apply the Floquet formalism to a periodically driven system, represented by the 
Hamiltonian  described by Eq.~(\ref{eq:tth}), which satisfies 
$\hat{\mathcal{H}}_\textbf{k}$(t) = $\hat{\mathcal{H}}_\textbf{k}(t + \tau_{\textrm{ph}})$, where $\tau_{\textrm{ph}}$ is the time-period corresponds to $\omega_{\textrm{ph}}$. 
The Hamiltonian obeys the time-dependent Schr\"odinger equation and its  solution is  
obtained in the basis of the Floquet states as $|\psi_n^{\rm F}(t)\rangle = e^{-i\epsilon_n^{\rm F}t}|\phi_n^{\rm F}(t)\rangle$. 
Here, $\epsilon_n^{\rm F}$ is the quasi-energy corresponds to the $n^{\rm th}$ Floquet state, 
and  $|\phi_n^{\rm F}(t)\rangle$ is the time-periodic part of the wave function, such that $|\phi_n^{\rm F}(t+\tau_{\textrm{ph}})\rangle = |\phi_n^{\rm F}(t)\rangle$. 
The DSs in a Floquet system are the combined spatio-temporal symmetries, which provide different kinds of  selection rules as discussed in Ref.~\cite{neufeld2019floquet,nagai2020dynamical}. 

In the presence of the probe pulse, 
the laser-graphene 
interaction within tight-binding approximation can be modelled with the Peierls substitution as 
$\hat{\mathcal{H}}_{\textbf{k}}(t) \rightarrow \hat{\mathcal{H}}_{\textbf{k}+\textbf{A}(t)}(t)$.
For the sake of simplicity, we employ  a perturbative approach to understand the polarisation of the sidebands as the strength of the sidebands is much weaker in comparison 
to the main harmonic peaks. 
Let us expand $\hat{\mathcal{H}}_{\textbf{k}+\textbf{A}(t)}(t)$ in terms of $i\textbf{A}(t)\cdot \textbf{d}_i(t)$ as 
\begin{equation}
	\hat{\mathcal{H}}_{\textbf{k}+\textbf{A}(t)}(t) \approx \hat{\mathcal{H}}_{\textbf{k}}(t) + \textbf{A}(t)\cdot \nabla_\textbf{k} \hat{\mathcal{H}}_{\textbf{k}}(t). \label{eq:h_approx}
\end{equation}
The second term in the above equation can be treated as perturbation as 
 $\hat{\mathcal{H}}_{\textbf{k}}^\prime(t) = \textbf{A}(t)\cdot\hat{\textbf{J}}(t)$ with 
$\hat{\textbf{J}} = \nabla_{\textbf{A}(t)}\hat{\mathcal{H}}_{\textbf{k}+\textbf{A}(t)}$ 
is the current operator in the Bloch basis. In Eq.~(\ref{eq:h_approx}), 
higher-order terms are neglected.
 
By following Ref.~\cite{nagai2020dynamical} and 
assuming the electron initially is in the Floquet state $|\phi_i^{\rm F}\rangle$, we can  
solve time-dependent Schr\"odinger equation within first-order perturbation theory and 
the $\mu^{\rm th}$-component of the current can be written as 
\begin{equation}
	\begin{split}
	   J_\mu(t) &= \langle \phi_i^{\rm F}(t)| \hat{J}_\mu(t) | \phi_i^{\rm F}(t)\rangle \\
		&-\sum_{e\neq i} \int_{-\infty}^t  i dt' e^{-i\omega_{ei}(t-t')}\chi^{\rm F}_{\mu\nu}(t,t') A_\nu(t') 
		+ \textrm{c.c.} 
	\end{split}
\end{equation}
Here, $\chi^{\rm F}_{\mu\nu}(t,t') = \langle \phi_i^{\rm F}(t)| \hat{J}_\mu(t) | \phi_e^{\rm F}(t)\rangle \langle \phi_e^{\rm F}(t')| \hat{J}_\nu(t') | \phi_i^{\rm F}(t')\rangle$. 
From the above equation, it is apparent that the second term correlates to the generations of the sidebands via  Raman process.  

The symmetry constraint for the $m^{\rm th}$-order sideband can be written as 
$\hat{X}^t\textbf{E}_{s,m}(t) [\hat{X}^t\textbf{E}(t)]^\dagger$ = $\textbf{E}_{s,m} \textbf{E}^\dagger(t)$, provided spatial symmetries of $\hat{X}^t$ and probe-pulse are same~\cite{nagai2020dynamical}. 
Here, $\textbf{E}_{s,m}(t)$ and $\textbf{E}(t)$ are, respectively, the electric fields associated with 
$m^{\rm th}$-order sideband and the probe laser; and $\hat{X}^t$ is the dynamical symmetry operation. 
The quantity $\textbf{E}_{s,m}(t)\textbf{E}(t)^\dagger$ is denoted by  $\mathcal{R}_m(t)$ and known as Raman tensor ~\cite{nagai2020dynamical}. Thus the selection rules for the sidebands depend on the invariance of the Raman tensor under operation with the DSs of the Floquet system.

\begin{figure}
\includegraphics[width=\linewidth]{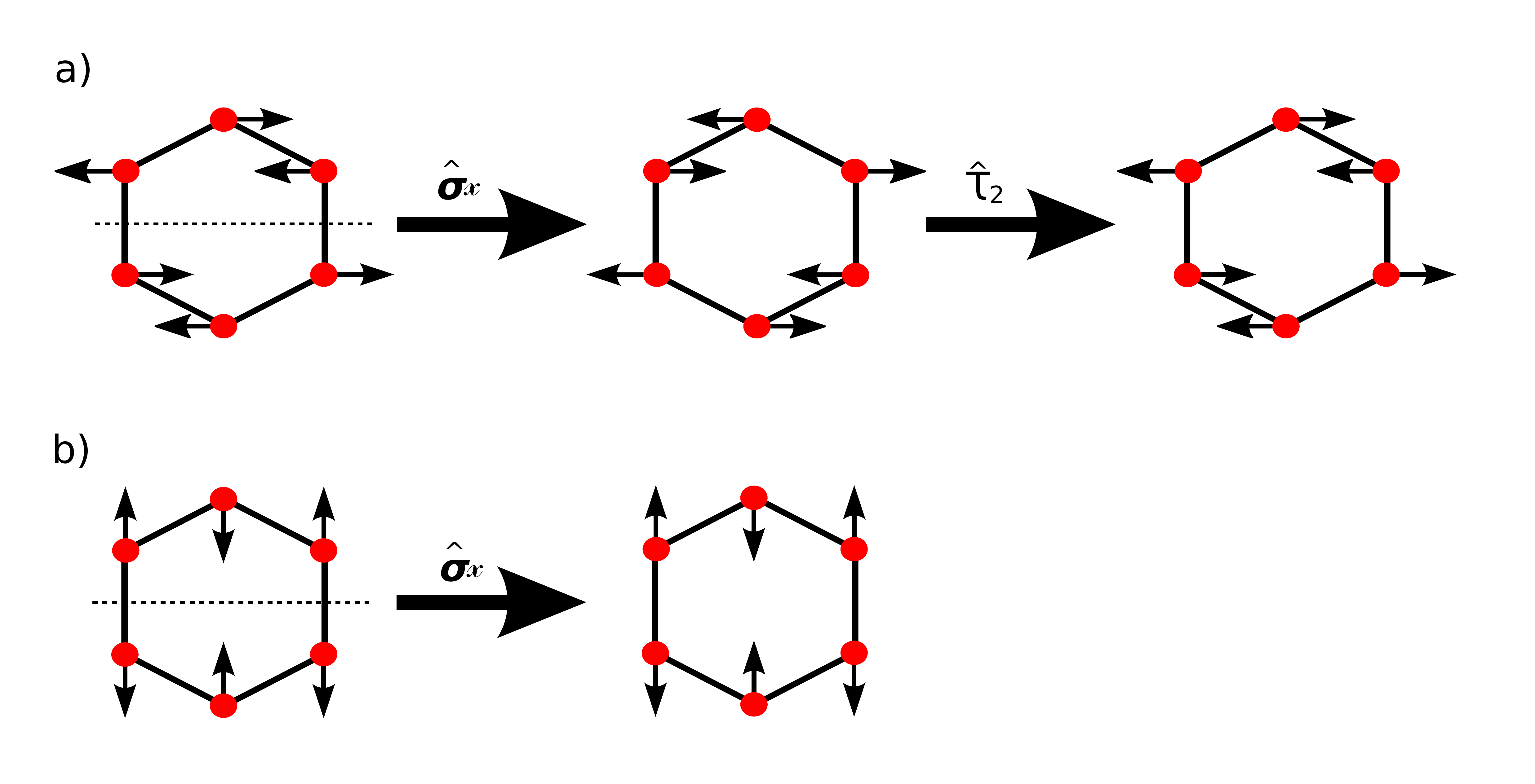}
\caption{Schematic representations of the dynamical symmetries of the Floquet Hamiltonian 
(a) $\hat{\mathcal{D}}_{1}$ = $\hat{\sigma_{x}}\cdot \hat{\tau}_2$ 
(b) $\hat{\mathcal{D}}_{2}$ = $\hat{\sigma_{x}}$. The arrows show the displacements of the atom for a particular phonon mode.}
\label{DS}
\end{figure}

There are two DSs corresponding to the coherent $\textsf{iLO}$ phonon mode as 
shown in Fig.~\ref{DS}. 
We define $\tau_n$ as the time translation of $\tau_{\textrm{ph}}/n$, $\hat{C}_{n\mu}$ is the 
rotation of 2$\pi/n$ with respect to $\mu$-axis, $\hat{\sigma}_\mu$ is the reflection with respect to 
$\mu$-axis, and $\hat{\mathcal{T}}$ is the time-reversal operator. The symmetry operations 
$\mathcal{D}_1 = \hat{\sigma}_x \cdot \tau_2$ [see Fig.~\ref{DS}(a)], and $\mathcal{D}_2 = \hat{\sigma_{x}}$ [see Fig.~\ref{DS}(b)] leaves the system invariant.

The selection rules for the sidebands and its polarisation directions are obtained from the DSs as 
shown in Fig.~\ref{DS} and requires  a condition as 
$\hat{\mathcal{D}}\mathcal{R}_m(t) = \mathcal{R}_m(t)$.
We assume that the temporal part of the  $m^{\rm th}$-order sideband as 
$e^{i(\omega_0 \pm  m\omega_{\textrm{ph}})t+\phi_0}$, 
and of the probe laser pulse as $e^{i\omega_0 t}$. 
In such situation, the Raman tensor is explicitly written as
\begin{equation}
	\mathcal{R}_m(t) = e^{i(\pm m \omega_{\textrm{ph}}t + \phi_0)} \begin{bmatrix}
		E_{s,m_{x}}E^{*}_{x} & E_{s,m_{x}}E^{*}_{y} \\
		E_{s,m_{y}}E^{*}_{x} & E_{s,m_{y}}E^{*}_{y}
	\end{bmatrix}.
\end{equation}

When the probe laser is polarised along \textsf{X}-axis, the invariance condition for the 
Raman tensor $\hat{\mathcal{D}_1}\mathcal{R}_m(t) = \mathcal{R}_m(t)$ reduces to 
\begin{equation}
	e^{i(\pm m\omega_{\textrm{ph}}t)} \begin{bmatrix}
		E_{s,m_{x}}  \\
		E_{s,m_{y}}
	\end{bmatrix}
	= e^{i[\pm m(\omega_{\textrm{ph}}t +\pi)]} \begin{bmatrix}
		E_{s,m_{x}}  \\
		-E_{s,m_{y}}
	\end{bmatrix}.
\label{selm1}
\end{equation}

The selection rule for the $m^{\rm th}$-order sideband is as follows:  
when $m$ is odd (even), the polarisation of the sideband 
will be along the \textsf{Y}(\textsf{X}) direction. 
Our observations in Fig.~$\ref{HHGlattice}$(a) are consistent with 
Eq.~(\ref{selm1}).

When the $\textsf{iLO}$  phonon mode is excited and the probe pulse 
is along  $\mathsf{\Gamma-M}$ direction,  
$ \hat{\sigma}_y \cdot \tau_2$, and $\hat{C}_{2Z}$ are the DSs, which leave the 
Raman tensor invariant. 
It is straightforward to see that the  selection rules for the $m^{\rm th}$-order sideband are deduced as: 
when $m$  
is odd (even), the polarisation of the sidebands will be along the \textsf{X}(\textsf{Y}) direction. 
On the other hand, when  the $\textsf{iTO}$  phonon mode is excited and the probe pulse
is along $\mathsf{\Gamma-K}$ ($\mathsf{\Gamma-M}$) direction, 
$\hat{\sigma}_x$ ($\hat{\sigma}_y$) is the DS, which yields Raman tensor  invariant [see Fig.~\ref{DS}(b)]. 
This symmetry restricts the polarisation of the sidebands to be along the direction of the probe pulse. 
Our results are consistent with the observation made 
in Fig. $\ref{HHGlattice}$.
With the increased intensity of the probe, higher-order harmonics and sidebands will appear. 

To summarise, we have established that high-harmonic spectroscopy is responsive to the coherent lattice dynamics in solids. 
The  high-harmonic spectrum is modulated by the frequency of the excited phonon mode within the solid. Both  in-plane E$_{2g}$ Raman-active phonon modes of the monolayer graphene lead to the generation of higher-order sidebands, along with the main harmonic peaks. 
In the case of  $\textsf{iLO}$  phonon mode excitation, the even- and odd-order sidebands are 
polarised parallel and perpendicular to the polarisation of probe harmonic pulse, respectively.  
In the case of  $\textsf{iTO}$  phonon mode, 
all sidebands are polarised along the probe harmonic pulse's polarisation. 
The polarisations of the sidebands are 
dictated by the dynamical symmetries of the combined system, which includes the phonon modes and probe laser pulse. Therefore, the polarisation properties  are 
a sensitive probe of these dynamical symmetries.  
The presence of high-harmonic signal perpendicular to the 
polarisation of the probe pulse is a signature of  
lattice excitation-driven 
symmetry breaking of the reflection plane. 
The present work is paving a way  for probing phonon-driven processes in solids and 
non-linear phononics with sub-cycle temporal resolution.

\section*{Acknowledgements}
We acknowledge fruitful discussion with Sumiran Pujari (IIT Bombay), Dipanshu Bansal (IIT Bombay) and Klaus Reimann (MBI Berlin).  
G. D. acknowledges support from Science and Engineering Research Board (SERB) India 
(Project No. MTR/2021/000138).  
D.K. acknowledges support from CRC 1375 ``NOA--Nonlinear optics down
to atomic scales", Project C4, funded by the Deutsche Forschungsgemeinschaft (DFG).

%\newpage

%\bibliography{solid_HHG}

\begin{thebibliography}{66}%
\makeatletter
\providecommand \@ifxundefined [1]{%
 \@ifx{#1\undefined}
}%
\providecommand \@ifnum [1]{%
 \ifnum #1\expandafter \@firstoftwo
 \else \expandafter \@secondoftwo
 \fi
}%
\providecommand \@ifx [1]{%
 \ifx #1\expandafter \@firstoftwo
 \else \expandafter \@secondoftwo
 \fi
}%
\providecommand \natexlab [1]{#1}%
\providecommand \enquote  [1]{``#1''}%
\providecommand \bibnamefont  [1]{#1}%
\providecommand \bibfnamefont [1]{#1}%
\providecommand \citenamefont [1]{#1}%
\providecommand \href@noop [0]{\@secondoftwo}%
\providecommand \href [0]{\begingroup \@sanitize@url \@href}%
\providecommand \@href[1]{\@@startlink{#1}\@@href}%
\providecommand \@@href[1]{\endgroup#1\@@endlink}%
\providecommand \@sanitize@url [0]{\catcode `\\12\catcode `\$12\catcode
  `\&12\catcode `\#12\catcode `\^12\catcode `\_12\catcode `\%12\relax}%
\providecommand \@@startlink[1]{}%
\providecommand \@@endlink[0]{}%
\providecommand \url  [0]{\begingroup\@sanitize@url \@url }%
\providecommand \@url [1]{\endgroup\@href {#1}{\urlprefix }}%
\providecommand \urlprefix  [0]{URL }%
\providecommand \Eprint [0]{\href }%
\providecommand \doibase [0]{http://dx.doi.org/}%
\providecommand \selectlanguage [0]{\@gobble}%
\providecommand \bibinfo  [0]{\@secondoftwo}%
\providecommand \bibfield  [0]{\@secondoftwo}%
\providecommand \translation [1]{[#1]}%
\providecommand \BibitemOpen [0]{}%
\providecommand \bibitemStop [0]{}%
\providecommand \bibitemNoStop [0]{.\EOS\space}%
\providecommand \EOS [0]{\spacefactor3000\relax}%
\providecommand \BibitemShut  [1]{\csname bibitem#1\endcsname}%
\let\auto@bib@innerbib\@empty
%</preamble>
\bibitem [{\citenamefont {Ferray}\ \emph {et~al.}(1988)\citenamefont {Ferray},
  \citenamefont {L'Huillier}, \citenamefont {Li}, \citenamefont {Lompre},
  \citenamefont {Mainfray},\ and\ \citenamefont {Manus}}]{ferray1988multiple}%
  \BibitemOpen
  \bibfield  {author} {\bibinfo {author} {\bibfnamefont {M.}~\bibnamefont
  {Ferray}}, \bibinfo {author} {\bibfnamefont {A.}~\bibnamefont {L'Huillier}},
  \bibinfo {author} {\bibfnamefont {X.~F.}\ \bibnamefont {Li}}, \bibinfo
  {author} {\bibfnamefont {L.~A.}\ \bibnamefont {Lompre}}, \bibinfo {author}
  {\bibfnamefont {G.}~\bibnamefont {Mainfray}}, \ and\ \bibinfo {author}
  {\bibfnamefont {C.}~\bibnamefont {Manus}},\ }\href@noop {} {\bibfield
  {journal} {\bibinfo  {journal} {Journal of Physics B}\ }\textbf {\bibinfo
  {volume} {21}},\ \bibinfo {pages} {L31} (\bibinfo {year} {1988})}\BibitemShut
  {NoStop}%
\bibitem [{\citenamefont {Ghimire}\ \emph {et~al.}(2011)\citenamefont
  {Ghimire}, \citenamefont {DiChiara}, \citenamefont {Sistrunk}, \citenamefont
  {Agostini}, \citenamefont {DiMauro},\ and\ \citenamefont
  {Reis}}]{ghimire2011observation}%
  \BibitemOpen
  \bibfield  {author} {\bibinfo {author} {\bibfnamefont {S.}~\bibnamefont
  {Ghimire}}, \bibinfo {author} {\bibfnamefont {A.~D.}\ \bibnamefont
  {DiChiara}}, \bibinfo {author} {\bibfnamefont {E.}~\bibnamefont {Sistrunk}},
  \bibinfo {author} {\bibfnamefont {P.}~\bibnamefont {Agostini}}, \bibinfo
  {author} {\bibfnamefont {L.~F.}\ \bibnamefont {DiMauro}}, \ and\ \bibinfo
  {author} {\bibfnamefont {D.~A.}\ \bibnamefont {Reis}},\ }\href@noop {}
  {\bibfield  {journal} {\bibinfo  {journal} {Nature Physics}\ }\textbf
  {\bibinfo {volume} {7}},\ \bibinfo {pages} {138} (\bibinfo {year}
  {2011})}\BibitemShut {NoStop}%
\bibitem [{\citenamefont {Vampa}\ \emph {et~al.}(2015)\citenamefont {Vampa},
  \citenamefont {Hammond}, \citenamefont {Thir{\'e}}, \citenamefont {Schmidt},
  \citenamefont {L{\'e}gar{\'e}}, \citenamefont {McDonald}, \citenamefont
  {Brabec}, \citenamefont {Klug},\ and\ \citenamefont {Corkum}}]{vampa2015all}%
  \BibitemOpen
  \bibfield  {author} {\bibinfo {author} {\bibfnamefont {G.}~\bibnamefont
  {Vampa}}, \bibinfo {author} {\bibfnamefont {T.~J.}\ \bibnamefont {Hammond}},
  \bibinfo {author} {\bibfnamefont {N.}~\bibnamefont {Thir{\'e}}}, \bibinfo
  {author} {\bibfnamefont {B.~E.}\ \bibnamefont {Schmidt}}, \bibinfo {author}
  {\bibfnamefont {F.}~\bibnamefont {L{\'e}gar{\'e}}}, \bibinfo {author}
  {\bibfnamefont {C.~R.}\ \bibnamefont {McDonald}}, \bibinfo {author}
  {\bibfnamefont {T.}~\bibnamefont {Brabec}}, \bibinfo {author} {\bibfnamefont
  {D.~D.}\ \bibnamefont {Klug}}, \ and\ \bibinfo {author} {\bibfnamefont
  {P.~B.}\ \bibnamefont {Corkum}},\ }\href@noop {} {\bibfield  {journal}
  {\bibinfo  {journal} {Physical Review Letters}\ }\textbf {\bibinfo {volume}
  {115}},\ \bibinfo {pages} {193603} (\bibinfo {year} {2015})}\BibitemShut
  {NoStop}%
\bibitem [{\citenamefont {Luu}\ \emph {et~al.}(2015)\citenamefont {Luu},
  \citenamefont {Garg}, \citenamefont {Kruchinin}, \citenamefont {Moulet},
  \citenamefont {Hassan},\ and\ \citenamefont {Goulielmakis}}]{luu2015extreme}%
  \BibitemOpen
  \bibfield  {author} {\bibinfo {author} {\bibfnamefont {T.~T.}\ \bibnamefont
  {Luu}}, \bibinfo {author} {\bibfnamefont {M.}~\bibnamefont {Garg}}, \bibinfo
  {author} {\bibfnamefont {S.~Y.}\ \bibnamefont {Kruchinin}}, \bibinfo {author}
  {\bibfnamefont {A.}~\bibnamefont {Moulet}}, \bibinfo {author} {\bibfnamefont
  {M.~T.}\ \bibnamefont {Hassan}}, \ and\ \bibinfo {author} {\bibfnamefont
  {E.}~\bibnamefont {Goulielmakis}},\ }\href@noop {} {\bibfield  {journal}
  {\bibinfo  {journal} {Nature}\ }\textbf {\bibinfo {volume} {521}},\ \bibinfo
  {pages} {498} (\bibinfo {year} {2015})}\BibitemShut {NoStop}%
\bibitem [{\citenamefont {Lanin}\ \emph {et~al.}(2017)\citenamefont {Lanin},
  \citenamefont {Stepanov}, \citenamefont {Fedotov},\ and\ \citenamefont
  {Zheltikov}}]{lanin2017mapping}%
  \BibitemOpen
  \bibfield  {author} {\bibinfo {author} {\bibfnamefont {A.~A.}\ \bibnamefont
  {Lanin}}, \bibinfo {author} {\bibfnamefont {E.~A.}\ \bibnamefont {Stepanov}},
  \bibinfo {author} {\bibfnamefont {A.~B.}\ \bibnamefont {Fedotov}}, \ and\
  \bibinfo {author} {\bibfnamefont {A.~M.}\ \bibnamefont {Zheltikov}},\
  }\href@noop {} {\bibfield  {journal} {\bibinfo  {journal} {Optica}\ }\textbf
  {\bibinfo {volume} {4}},\ \bibinfo {pages} {516} (\bibinfo {year}
  {2017})}\BibitemShut {NoStop}%
\bibitem [{\citenamefont {Mrudul}\ \emph {et~al.}(2019)\citenamefont {Mrudul},
  \citenamefont {Pattanayak}, \citenamefont {Ivanov},\ and\ \citenamefont
  {Dixit}}]{pattanayak2019direct}%
  \BibitemOpen
  \bibfield  {author} {\bibinfo {author} {\bibfnamefont {M.~S.}\ \bibnamefont
  {Mrudul}}, \bibinfo {author} {\bibfnamefont {A.}~\bibnamefont {Pattanayak}},
  \bibinfo {author} {\bibfnamefont {M.}~\bibnamefont {Ivanov}}, \ and\ \bibinfo
  {author} {\bibfnamefont {G.}~\bibnamefont {Dixit}},\ }\href@noop {}
  {\bibfield  {journal} {\bibinfo  {journal} {Physical Review A}\ }\textbf
  {\bibinfo {volume} {100}},\ \bibinfo {pages} {043420} (\bibinfo {year}
  {2019})}\BibitemShut {NoStop}%
\bibitem [{\citenamefont {Tancogne-Dejean}\ \emph {et~al.}(2017)\citenamefont
  {Tancogne-Dejean}, \citenamefont {M{\"u}cke}, \citenamefont {K{\"a}rtner},\
  and\ \citenamefont {Rubio}}]{tancogne2017impact}%
  \BibitemOpen
  \bibfield  {author} {\bibinfo {author} {\bibfnamefont {N.}~\bibnamefont
  {Tancogne-Dejean}}, \bibinfo {author} {\bibfnamefont {O.~D.}\ \bibnamefont
  {M{\"u}cke}}, \bibinfo {author} {\bibfnamefont {F.~X.}\ \bibnamefont
  {K{\"a}rtner}}, \ and\ \bibinfo {author} {\bibfnamefont {A.}~\bibnamefont
  {Rubio}},\ }\href@noop {} {\bibfield  {journal} {\bibinfo  {journal}
  {Physical review letters}\ }\textbf {\bibinfo {volume} {118}},\ \bibinfo
  {pages} {087403} (\bibinfo {year} {2017})}\BibitemShut {NoStop}%
\bibitem [{\citenamefont {Mrudul}\ \emph {et~al.}(2020)\citenamefont {Mrudul},
  \citenamefont {Tancogne-Dejean}, \citenamefont {Rubio},\ and\ \citenamefont
  {Dixit}}]{mrudul2020high}%
  \BibitemOpen
  \bibfield  {author} {\bibinfo {author} {\bibfnamefont {M.~S.}\ \bibnamefont
  {Mrudul}}, \bibinfo {author} {\bibfnamefont {N.}~\bibnamefont
  {Tancogne-Dejean}}, \bibinfo {author} {\bibfnamefont {A.}~\bibnamefont
  {Rubio}}, \ and\ \bibinfo {author} {\bibfnamefont {G.}~\bibnamefont
  {Dixit}},\ }\href@noop {} {\bibfield  {journal} {\bibinfo  {journal} {npj
  Computational Materials}\ }\textbf {\bibinfo {volume} {6}},\ \bibinfo {pages}
  {1} (\bibinfo {year} {2020})}\BibitemShut {NoStop}%
\bibitem [{\citenamefont {Pattanayak}\ \emph {et~al.}(2020)\citenamefont
  {Pattanayak}, \citenamefont {Mrudul},\ and\ \citenamefont
  {Dixit}}]{pattanayak2020influence}%
  \BibitemOpen
  \bibfield  {author} {\bibinfo {author} {\bibfnamefont {A.}~\bibnamefont
  {Pattanayak}}, \bibinfo {author} {\bibfnamefont {M.~S.}\ \bibnamefont
  {Mrudul}}, \ and\ \bibinfo {author} {\bibfnamefont {G.}~\bibnamefont
  {Dixit}},\ }\href@noop {} {\bibfield  {journal} {\bibinfo  {journal}
  {Physical Review A}\ }\textbf {\bibinfo {volume} {101}},\ \bibinfo {pages}
  {013404} (\bibinfo {year} {2020})}\BibitemShut {NoStop}%
\bibitem [{\citenamefont {Mrudul}\ \emph {et~al.}(2021)\citenamefont {Mrudul},
  \citenamefont {Jim{\'e}nez-Gal{\'a}n}, \citenamefont {Ivanov},\ and\
  \citenamefont {Dixit}}]{mrudul2021light}%
  \BibitemOpen
  \bibfield  {author} {\bibinfo {author} {\bibfnamefont {M.}~\bibnamefont
  {Mrudul}}, \bibinfo {author} {\bibfnamefont {{\'A}.}~\bibnamefont
  {Jim{\'e}nez-Gal{\'a}n}}, \bibinfo {author} {\bibfnamefont {M.}~\bibnamefont
  {Ivanov}}, \ and\ \bibinfo {author} {\bibfnamefont {G.}~\bibnamefont
  {Dixit}},\ }\href@noop {} {\bibfield  {journal} {\bibinfo  {journal}
  {Optica}\ }\textbf {\bibinfo {volume} {8}},\ \bibinfo {pages} {422} (\bibinfo
  {year} {2021})}\BibitemShut {NoStop}%
\bibitem [{\citenamefont {Jim{\'e}nez-Gal{\'a}n}\ \emph
  {et~al.}(2020)\citenamefont {Jim{\'e}nez-Gal{\'a}n}, \citenamefont {Silva},
  \citenamefont {Smirnova},\ and\ \citenamefont
  {Ivanov}}]{jimenez2020lightwave}%
  \BibitemOpen
  \bibfield  {author} {\bibinfo {author} {\bibfnamefont {{\'A}.}~\bibnamefont
  {Jim{\'e}nez-Gal{\'a}n}}, \bibinfo {author} {\bibfnamefont {R.}~\bibnamefont
  {Silva}}, \bibinfo {author} {\bibfnamefont {O.}~\bibnamefont {Smirnova}}, \
  and\ \bibinfo {author} {\bibfnamefont {M.}~\bibnamefont {Ivanov}},\
  }\href@noop {} {\bibfield  {journal} {\bibinfo  {journal} {Nature Photonics}\
  }\textbf {\bibinfo {volume} {14}},\ \bibinfo {pages} {728} (\bibinfo {year}
  {2020})}\BibitemShut {NoStop}%
\bibitem [{\citenamefont {Langer}\ \emph {et~al.}(2018)\citenamefont {Langer},
  \citenamefont {Schmid}, \citenamefont {Schlauderer}, \citenamefont {Gmitra},
  \citenamefont {Fabian}, \citenamefont {Nagler}, \citenamefont {Sch{\"u}ller},
  \citenamefont {Korn}, \citenamefont {Hawkins}, \citenamefont {Steiner} \emph
  {et~al.}}]{langer2018lightwave}%
  \BibitemOpen
  \bibfield  {author} {\bibinfo {author} {\bibfnamefont {F.}~\bibnamefont
  {Langer}}, \bibinfo {author} {\bibfnamefont {C.~P.}\ \bibnamefont {Schmid}},
  \bibinfo {author} {\bibfnamefont {S.}~\bibnamefont {Schlauderer}}, \bibinfo
  {author} {\bibfnamefont {M.}~\bibnamefont {Gmitra}}, \bibinfo {author}
  {\bibfnamefont {J.}~\bibnamefont {Fabian}}, \bibinfo {author} {\bibfnamefont
  {P.}~\bibnamefont {Nagler}}, \bibinfo {author} {\bibfnamefont
  {C.}~\bibnamefont {Sch{\"u}ller}}, \bibinfo {author} {\bibfnamefont
  {T.}~\bibnamefont {Korn}}, \bibinfo {author} {\bibfnamefont {P.}~\bibnamefont
  {Hawkins}}, \bibinfo {author} {\bibfnamefont {J.}~\bibnamefont {Steiner}},
  \emph {et~al.},\ }\href@noop {} {\bibfield  {journal} {\bibinfo  {journal}
  {Nature}\ }\textbf {\bibinfo {volume} {557}},\ \bibinfo {pages} {76}
  (\bibinfo {year} {2018})}\BibitemShut {NoStop}%
\bibitem [{\citenamefont {Mrudul}\ and\ \citenamefont
  {Dixit}(2021{\natexlab{a}})}]{mrudul2021controlling}%
  \BibitemOpen
  \bibfield  {author} {\bibinfo {author} {\bibfnamefont {M.}~\bibnamefont
  {Mrudul}}\ and\ \bibinfo {author} {\bibfnamefont {G.}~\bibnamefont {Dixit}},\
  }\href@noop {} {\bibfield  {journal} {\bibinfo  {journal} {Journal of Physics
  B}\ }\textbf {\bibinfo {volume} {54}},\ \bibinfo {pages} {224001} (\bibinfo
  {year} {2021}{\natexlab{a}})}\BibitemShut {NoStop}%
\bibitem [{\citenamefont {Schubert}\ \emph {et~al.}(2014)\citenamefont
  {Schubert}, \citenamefont {Hohenleutner}, \citenamefont {Langer},
  \citenamefont {Urbanek}, \citenamefont {Lange}, \citenamefont {Huttner},
  \citenamefont {Golde}, \citenamefont {Meier}, \citenamefont {Kira},
  \citenamefont {Koch},\ and\ \citenamefont {Huber}}]{schubert2014sub}%
  \BibitemOpen
  \bibfield  {author} {\bibinfo {author} {\bibfnamefont {O.}~\bibnamefont
  {Schubert}}, \bibinfo {author} {\bibfnamefont {M.}~\bibnamefont
  {Hohenleutner}}, \bibinfo {author} {\bibfnamefont {F.}~\bibnamefont
  {Langer}}, \bibinfo {author} {\bibfnamefont {B.}~\bibnamefont {Urbanek}},
  \bibinfo {author} {\bibfnamefont {C.}~\bibnamefont {Lange}}, \bibinfo
  {author} {\bibfnamefont {U.}~\bibnamefont {Huttner}}, \bibinfo {author}
  {\bibfnamefont {D.}~\bibnamefont {Golde}}, \bibinfo {author} {\bibfnamefont
  {T.}~\bibnamefont {Meier}}, \bibinfo {author} {\bibfnamefont
  {M.}~\bibnamefont {Kira}}, \bibinfo {author} {\bibfnamefont {S.~W.}\
  \bibnamefont {Koch}}, \ and\ \bibinfo {author} {\bibfnamefont
  {R.}~\bibnamefont {Huber}},\ }\href@noop {} {\bibfield  {journal} {\bibinfo
  {journal} {Nature Photonics}\ }\textbf {\bibinfo {volume} {8}},\ \bibinfo
  {pages} {119} (\bibinfo {year} {2014})}\BibitemShut {NoStop}%
\bibitem [{\citenamefont {Bauer}\ and\ \citenamefont
  {Hansen}(2018)}]{bauer2018high}%
  \BibitemOpen
  \bibfield  {author} {\bibinfo {author} {\bibfnamefont {D.}~\bibnamefont
  {Bauer}}\ and\ \bibinfo {author} {\bibfnamefont {K.~K.}\ \bibnamefont
  {Hansen}},\ }\href@noop {} {\bibfield  {journal} {\bibinfo  {journal}
  {Physical Review Letters}\ }\textbf {\bibinfo {volume} {120}},\ \bibinfo
  {pages} {177401} (\bibinfo {year} {2018})}\BibitemShut {NoStop}%
\bibitem [{\citenamefont {Silva}\ \emph {et~al.}(2018)\citenamefont {Silva},
  \citenamefont {Blinov}, \citenamefont {Rubtsov}, \citenamefont {Smirnova},\
  and\ \citenamefont {Ivanov}}]{silva2018high}%
  \BibitemOpen
  \bibfield  {author} {\bibinfo {author} {\bibfnamefont {R.~E.~F.}\
  \bibnamefont {Silva}}, \bibinfo {author} {\bibfnamefont {I.~V.}\ \bibnamefont
  {Blinov}}, \bibinfo {author} {\bibfnamefont {A.~N.}\ \bibnamefont {Rubtsov}},
  \bibinfo {author} {\bibfnamefont {O.}~\bibnamefont {Smirnova}}, \ and\
  \bibinfo {author} {\bibfnamefont {M.}~\bibnamefont {Ivanov}},\ }\href@noop {}
  {\bibfield  {journal} {\bibinfo  {journal} {Nature Photonics}\ }\textbf
  {\bibinfo {volume} {12}},\ \bibinfo {pages} {266} (\bibinfo {year}
  {2018})}\BibitemShut {NoStop}%
\bibitem [{\citenamefont {Bai}\ \emph {et~al.}(2020)\citenamefont {Bai},
  \citenamefont {Fei}, \citenamefont {Wang}, \citenamefont {Li}, \citenamefont
  {Li}, \citenamefont {Song}, \citenamefont {Li}, \citenamefont {Xu},\ and\
  \citenamefont {Liu}}]{bai2020high}%
  \BibitemOpen
  \bibfield  {author} {\bibinfo {author} {\bibfnamefont {Y.}~\bibnamefont
  {Bai}}, \bibinfo {author} {\bibfnamefont {F.}~\bibnamefont {Fei}}, \bibinfo
  {author} {\bibfnamefont {S.}~\bibnamefont {Wang}}, \bibinfo {author}
  {\bibfnamefont {N.}~\bibnamefont {Li}}, \bibinfo {author} {\bibfnamefont
  {X.}~\bibnamefont {Li}}, \bibinfo {author} {\bibfnamefont {F.}~\bibnamefont
  {Song}}, \bibinfo {author} {\bibfnamefont {R.}~\bibnamefont {Li}}, \bibinfo
  {author} {\bibfnamefont {Z.}~\bibnamefont {Xu}}, \ and\ \bibinfo {author}
  {\bibfnamefont {P.}~\bibnamefont {Liu}},\ }\href@noop {} {\bibfield
  {journal} {\bibinfo  {journal} {Nature Physics}\ ,\ \bibinfo {pages} {1}}
  (\bibinfo {year} {2020})}\BibitemShut {NoStop}%
\bibitem [{\citenamefont {Imai}\ \emph {et~al.}(2020)\citenamefont {Imai},
  \citenamefont {Ono},\ and\ \citenamefont {Ishihara}}]{imai2020high}%
  \BibitemOpen
  \bibfield  {author} {\bibinfo {author} {\bibfnamefont {S.}~\bibnamefont
  {Imai}}, \bibinfo {author} {\bibfnamefont {A.}~\bibnamefont {Ono}}, \ and\
  \bibinfo {author} {\bibfnamefont {S.}~\bibnamefont {Ishihara}},\ }\href@noop
  {} {\bibfield  {journal} {\bibinfo  {journal} {Physical Review Letters}\
  }\textbf {\bibinfo {volume} {124}},\ \bibinfo {pages} {157404} (\bibinfo
  {year} {2020})}\BibitemShut {NoStop}%
\bibitem [{\citenamefont {Borsch}\ \emph {et~al.}(2020)\citenamefont {Borsch},
  \citenamefont {Schmid}, \citenamefont {Weigl}, \citenamefont {Schlauderer},
  \citenamefont {Hofmann}, \citenamefont {Lange}, \citenamefont {Steiner},
  \citenamefont {Koch}, \citenamefont {Huber},\ and\ \citenamefont
  {Kira}}]{borsch2020super}%
  \BibitemOpen
  \bibfield  {author} {\bibinfo {author} {\bibfnamefont {M.}~\bibnamefont
  {Borsch}}, \bibinfo {author} {\bibfnamefont {C.~P.}\ \bibnamefont {Schmid}},
  \bibinfo {author} {\bibfnamefont {L.}~\bibnamefont {Weigl}}, \bibinfo
  {author} {\bibfnamefont {S.}~\bibnamefont {Schlauderer}}, \bibinfo {author}
  {\bibfnamefont {N.}~\bibnamefont {Hofmann}}, \bibinfo {author} {\bibfnamefont
  {C.}~\bibnamefont {Lange}}, \bibinfo {author} {\bibfnamefont {J.~T.}\
  \bibnamefont {Steiner}}, \bibinfo {author} {\bibfnamefont {S.~W.}\
  \bibnamefont {Koch}}, \bibinfo {author} {\bibfnamefont {R.}~\bibnamefont
  {Huber}}, \ and\ \bibinfo {author} {\bibfnamefont {M.}~\bibnamefont {Kira}},\
  }\href@noop {} {\bibfield  {journal} {\bibinfo  {journal} {Science}\ }\textbf
  {\bibinfo {volume} {370}},\ \bibinfo {pages} {1204} (\bibinfo {year}
  {2020})}\BibitemShut {NoStop}%
\bibitem [{\citenamefont {Baykusheva}\ \emph {et~al.}(2021)\citenamefont
  {Baykusheva}, \citenamefont {Chac{\'o}n}, \citenamefont {Lu}, \citenamefont
  {Bailey}, \citenamefont {Sobota}, \citenamefont {Soifer}, \citenamefont
  {Kirchmann}, \citenamefont {Rotundu}, \citenamefont {Uher}, \citenamefont
  {Heinz} \emph {et~al.}}]{baykusheva2021all}%
  \BibitemOpen
  \bibfield  {author} {\bibinfo {author} {\bibfnamefont {D.}~\bibnamefont
  {Baykusheva}}, \bibinfo {author} {\bibfnamefont {A.}~\bibnamefont
  {Chac{\'o}n}}, \bibinfo {author} {\bibfnamefont {J.}~\bibnamefont {Lu}},
  \bibinfo {author} {\bibfnamefont {T.~P.}\ \bibnamefont {Bailey}}, \bibinfo
  {author} {\bibfnamefont {J.~A.}\ \bibnamefont {Sobota}}, \bibinfo {author}
  {\bibfnamefont {H.}~\bibnamefont {Soifer}}, \bibinfo {author} {\bibfnamefont
  {P.~S.}\ \bibnamefont {Kirchmann}}, \bibinfo {author} {\bibfnamefont
  {C.}~\bibnamefont {Rotundu}}, \bibinfo {author} {\bibfnamefont
  {C.}~\bibnamefont {Uher}}, \bibinfo {author} {\bibfnamefont {T.~F.}\
  \bibnamefont {Heinz}},  \emph {et~al.},\ }\href@noop {} {\bibfield  {journal}
  {\bibinfo  {journal} {Nano Letters}\ }\textbf {\bibinfo {volume} {21}},\
  \bibinfo {pages} {8970} (\bibinfo {year} {2021})}\BibitemShut {NoStop}%
\bibitem [{\citenamefont {Bharti}\ \emph {et~al.}(2022)\citenamefont {Bharti},
  \citenamefont {Mrudul},\ and\ \citenamefont {Dixit}}]{bharti2022high}%
  \BibitemOpen
  \bibfield  {author} {\bibinfo {author} {\bibfnamefont {A.}~\bibnamefont
  {Bharti}}, \bibinfo {author} {\bibfnamefont {M.}~\bibnamefont {Mrudul}}, \
  and\ \bibinfo {author} {\bibfnamefont {G.}~\bibnamefont {Dixit}},\
  }\href@noop {} {\bibfield  {journal} {\bibinfo  {journal} {Physical Review
  B}\ }\textbf {\bibinfo {volume} {105}},\ \bibinfo {pages} {155140} (\bibinfo
  {year} {2022})}\BibitemShut {NoStop}%
\bibitem [{\citenamefont {Pattanayak}\ \emph {et~al.}(2022)\citenamefont
  {Pattanayak}, \citenamefont {Pujari},\ and\ \citenamefont
  {Dixit}}]{pattanayak2022role}%
  \BibitemOpen
  \bibfield  {author} {\bibinfo {author} {\bibfnamefont {A.}~\bibnamefont
  {Pattanayak}}, \bibinfo {author} {\bibfnamefont {S.}~\bibnamefont {Pujari}},
  \ and\ \bibinfo {author} {\bibfnamefont {G.}~\bibnamefont {Dixit}},\
  }\href@noop {} {\bibfield  {journal} {\bibinfo  {journal} {Scientific
  Reports}\ }\textbf {\bibinfo {volume} {12}},\ \bibinfo {pages} {1} (\bibinfo
  {year} {2022})}\BibitemShut {NoStop}%
\bibitem [{\citenamefont {Chac{\'o}n}\ \emph {et~al.}(2020)\citenamefont
  {Chac{\'o}n}, \citenamefont {Kim}, \citenamefont {Zhu}, \citenamefont
  {Kelly}, \citenamefont {Dauphin}, \citenamefont {Pisanty}, \citenamefont
  {Maxwell}, \citenamefont {Pic{\'o}n}, \citenamefont {Ciappina}, \citenamefont
  {Kim} \emph {et~al.}}]{chacon2020circular}%
  \BibitemOpen
  \bibfield  {author} {\bibinfo {author} {\bibfnamefont {A.}~\bibnamefont
  {Chac{\'o}n}}, \bibinfo {author} {\bibfnamefont {D.}~\bibnamefont {Kim}},
  \bibinfo {author} {\bibfnamefont {W.}~\bibnamefont {Zhu}}, \bibinfo {author}
  {\bibfnamefont {S.~P.}\ \bibnamefont {Kelly}}, \bibinfo {author}
  {\bibfnamefont {A.}~\bibnamefont {Dauphin}}, \bibinfo {author} {\bibfnamefont
  {E.}~\bibnamefont {Pisanty}}, \bibinfo {author} {\bibfnamefont {A.~S.}\
  \bibnamefont {Maxwell}}, \bibinfo {author} {\bibfnamefont {A.}~\bibnamefont
  {Pic{\'o}n}}, \bibinfo {author} {\bibfnamefont {M.~F.}\ \bibnamefont
  {Ciappina}}, \bibinfo {author} {\bibfnamefont {D.~E.}\ \bibnamefont {Kim}},
  \emph {et~al.},\ }\href@noop {} {\bibfield  {journal} {\bibinfo  {journal}
  {Physical Review B}\ }\textbf {\bibinfo {volume} {102}},\ \bibinfo {pages}
  {134115} (\bibinfo {year} {2020})}\BibitemShut {NoStop}%
\bibitem [{\citenamefont {Shao}\ \emph {et~al.}(2022)\citenamefont {Shao},
  \citenamefont {Lu}, \citenamefont {Zhang}, \citenamefont {Yu}, \citenamefont
  {Tohyama},\ and\ \citenamefont {Lu}}]{shao2022high}%
  \BibitemOpen
  \bibfield  {author} {\bibinfo {author} {\bibfnamefont {C.}~\bibnamefont
  {Shao}}, \bibinfo {author} {\bibfnamefont {H.}~\bibnamefont {Lu}}, \bibinfo
  {author} {\bibfnamefont {X.}~\bibnamefont {Zhang}}, \bibinfo {author}
  {\bibfnamefont {C.}~\bibnamefont {Yu}}, \bibinfo {author} {\bibfnamefont
  {T.}~\bibnamefont {Tohyama}}, \ and\ \bibinfo {author} {\bibfnamefont
  {R.}~\bibnamefont {Lu}},\ }\href@noop {} {\bibfield  {journal} {\bibinfo
  {journal} {Physical Review Letters}\ }\textbf {\bibinfo {volume} {128}},\
  \bibinfo {pages} {047401} (\bibinfo {year} {2022})}\BibitemShut {NoStop}%
\bibitem [{\citenamefont {Lakhotia}\ \emph {et~al.}(2020)\citenamefont
  {Lakhotia}, \citenamefont {Kim}, \citenamefont {Zhan}, \citenamefont {Hu},
  \citenamefont {Meng},\ and\ \citenamefont
  {Goulielmakis}}]{lakhotia2020laser}%
  \BibitemOpen
  \bibfield  {author} {\bibinfo {author} {\bibfnamefont {H.}~\bibnamefont
  {Lakhotia}}, \bibinfo {author} {\bibfnamefont {H.~Y.}\ \bibnamefont {Kim}},
  \bibinfo {author} {\bibfnamefont {M.}~\bibnamefont {Zhan}}, \bibinfo {author}
  {\bibfnamefont {S.}~\bibnamefont {Hu}}, \bibinfo {author} {\bibfnamefont
  {S.}~\bibnamefont {Meng}}, \ and\ \bibinfo {author} {\bibfnamefont
  {E.}~\bibnamefont {Goulielmakis}},\ }\href@noop {} {\bibfield  {journal}
  {\bibinfo  {journal} {Nature}\ }\textbf {\bibinfo {volume} {583}},\ \bibinfo
  {pages} {55} (\bibinfo {year} {2020})}\BibitemShut {NoStop}%
\bibitem [{\citenamefont {Mankowsky}\ \emph {et~al.}(2016)\citenamefont
  {Mankowsky}, \citenamefont {F{\"o}rst},\ and\ \citenamefont
  {Cavalleri}}]{mankowsky2016non}%
  \BibitemOpen
  \bibfield  {author} {\bibinfo {author} {\bibfnamefont {R.}~\bibnamefont
  {Mankowsky}}, \bibinfo {author} {\bibfnamefont {M.}~\bibnamefont
  {F{\"o}rst}}, \ and\ \bibinfo {author} {\bibfnamefont {A.}~\bibnamefont
  {Cavalleri}},\ }\href@noop {} {\bibfield  {journal} {\bibinfo  {journal}
  {Reports on Progress in Physics}\ }\textbf {\bibinfo {volume} {79}},\
  \bibinfo {pages} {064503} (\bibinfo {year} {2016})}\BibitemShut {NoStop}%
\bibitem [{\citenamefont {Hollinger}\ \emph {et~al.}(2019)\citenamefont
  {Hollinger}, \citenamefont {Shumakova}, \citenamefont {Pug{\v{z}}lys},
  \citenamefont {Baltu{\v{s}}ka}, \citenamefont {Khujanov}, \citenamefont
  {Spielmann},\ and\ \citenamefont {Kartashov}}]{hollinger2019high}%
  \BibitemOpen
  \bibfield  {author} {\bibinfo {author} {\bibfnamefont {R.}~\bibnamefont
  {Hollinger}}, \bibinfo {author} {\bibfnamefont {V.}~\bibnamefont
  {Shumakova}}, \bibinfo {author} {\bibfnamefont {A.}~\bibnamefont
  {Pug{\v{z}}lys}}, \bibinfo {author} {\bibfnamefont {A.}~\bibnamefont
  {Baltu{\v{s}}ka}}, \bibinfo {author} {\bibfnamefont {S.}~\bibnamefont
  {Khujanov}}, \bibinfo {author} {\bibfnamefont {C.}~\bibnamefont {Spielmann}},
  \ and\ \bibinfo {author} {\bibfnamefont {D.}~\bibnamefont {Kartashov}},\ }in\
  \href@noop {} {\emph {\bibinfo {booktitle} {EPJ Web of Conferences}}},\ Vol.\
  \bibinfo {volume} {205}\ (\bibinfo {organization} {EDP Sciences},\ \bibinfo
  {year} {2019})\ p.\ \bibinfo {pages} {02025}\BibitemShut {NoStop}%
\bibitem [{\citenamefont {Patchkovskii}(2009)}]{patchkovskii2009nuclear}%
  \BibitemOpen
  \bibfield  {author} {\bibinfo {author} {\bibfnamefont {S.}~\bibnamefont
  {Patchkovskii}},\ }\href@noop {} {\bibfield  {journal} {\bibinfo  {journal}
  {Physical Review Letters}\ }\textbf {\bibinfo {volume} {102}},\ \bibinfo
  {pages} {253602} (\bibinfo {year} {2009})}\BibitemShut {NoStop}%
\bibitem [{\citenamefont {Wagner}\ \emph {et~al.}(2006)\citenamefont {Wagner},
  \citenamefont {W{\"u}est}, \citenamefont {Christov}, \citenamefont
  {Popmintchev}, \citenamefont {Zhou}, \citenamefont {Murnane},\ and\
  \citenamefont {Kapteyn}}]{wagner2006monitoring}%
  \BibitemOpen
  \bibfield  {author} {\bibinfo {author} {\bibfnamefont {N.~L.}\ \bibnamefont
  {Wagner}}, \bibinfo {author} {\bibfnamefont {A.}~\bibnamefont {W{\"u}est}},
  \bibinfo {author} {\bibfnamefont {I.~P.}\ \bibnamefont {Christov}}, \bibinfo
  {author} {\bibfnamefont {T.}~\bibnamefont {Popmintchev}}, \bibinfo {author}
  {\bibfnamefont {X.}~\bibnamefont {Zhou}}, \bibinfo {author} {\bibfnamefont
  {M.~M.}\ \bibnamefont {Murnane}}, \ and\ \bibinfo {author} {\bibfnamefont
  {H.~C.}\ \bibnamefont {Kapteyn}},\ }\href@noop {} {\bibfield  {journal}
  {\bibinfo  {journal} {Proceedings of the National Academy of Sciences}\
  }\textbf {\bibinfo {volume} {103}},\ \bibinfo {pages} {13279} (\bibinfo
  {year} {2006})}\BibitemShut {NoStop}%
\bibitem [{\citenamefont {Le}\ \emph {et~al.}(2012)\citenamefont {Le},
  \citenamefont {Morishita}, \citenamefont {Lucchese},\ and\ \citenamefont
  {Lin}}]{le2012theory}%
  \BibitemOpen
  \bibfield  {author} {\bibinfo {author} {\bibfnamefont {A.-T.}\ \bibnamefont
  {Le}}, \bibinfo {author} {\bibfnamefont {T.}~\bibnamefont {Morishita}},
  \bibinfo {author} {\bibfnamefont {R.}~\bibnamefont {Lucchese}}, \ and\
  \bibinfo {author} {\bibfnamefont {C.~D.}\ \bibnamefont {Lin}},\ }\href@noop
  {} {\bibfield  {journal} {\bibinfo  {journal} {Physical Review Letters}\
  }\textbf {\bibinfo {volume} {109}},\ \bibinfo {pages} {203004} (\bibinfo
  {year} {2012})}\BibitemShut {NoStop}%
\bibitem [{\citenamefont {Baker}\ \emph {et~al.}(2006)\citenamefont {Baker},
  \citenamefont {Robinson}, \citenamefont {Haworth}, \citenamefont {Teng},
  \citenamefont {Smith}, \citenamefont {Chirila?}, \citenamefont {Lein},
  \citenamefont {Tisch},\ and\ \citenamefont {Marangos}}]{baker2006probing}%
  \BibitemOpen
  \bibfield  {author} {\bibinfo {author} {\bibfnamefont {S.}~\bibnamefont
  {Baker}}, \bibinfo {author} {\bibfnamefont {J.~S.}\ \bibnamefont {Robinson}},
  \bibinfo {author} {\bibfnamefont {C.}~\bibnamefont {Haworth}}, \bibinfo
  {author} {\bibfnamefont {H.}~\bibnamefont {Teng}}, \bibinfo {author}
  {\bibfnamefont {R.}~\bibnamefont {Smith}}, \bibinfo {author} {\bibfnamefont
  {C.~C.}\ \bibnamefont {Chirila?}}, \bibinfo {author} {\bibfnamefont
  {M.}~\bibnamefont {Lein}}, \bibinfo {author} {\bibfnamefont {J.}~\bibnamefont
  {Tisch}}, \ and\ \bibinfo {author} {\bibfnamefont {J.~P.}\ \bibnamefont
  {Marangos}},\ }\href@noop {} {\bibfield  {journal} {\bibinfo  {journal}
  {Science}\ }\textbf {\bibinfo {volume} {312}},\ \bibinfo {pages} {424}
  (\bibinfo {year} {2006})}\BibitemShut {NoStop}%
\bibitem [{\citenamefont {Lein}(2005)}]{lein2005attosecond}%
  \BibitemOpen
  \bibfield  {author} {\bibinfo {author} {\bibfnamefont {M.}~\bibnamefont
  {Lein}},\ }\href@noop {} {\bibfield  {journal} {\bibinfo  {journal} {Physical
  review letters}\ }\textbf {\bibinfo {volume} {94}},\ \bibinfo {pages}
  {053004} (\bibinfo {year} {2005})}\BibitemShut {NoStop}%
\bibitem [{\citenamefont {W{\"o}rner}\ \emph {et~al.}(2011)\citenamefont
  {W{\"o}rner}, \citenamefont {Bertrand}, \citenamefont {Fabre}, \citenamefont
  {Higuet}, \citenamefont {Ruf}, \citenamefont {Dubrouil}, \citenamefont
  {Patchkovskii}, \citenamefont {Spanner}, \citenamefont {Mairesse},
  \citenamefont {Blanchet} \emph {et~al.}}]{worner2011conical}%
  \BibitemOpen
  \bibfield  {author} {\bibinfo {author} {\bibfnamefont {H.~J.}\ \bibnamefont
  {W{\"o}rner}}, \bibinfo {author} {\bibfnamefont {J.~B.}\ \bibnamefont
  {Bertrand}}, \bibinfo {author} {\bibfnamefont {B.}~\bibnamefont {Fabre}},
  \bibinfo {author} {\bibfnamefont {J.}~\bibnamefont {Higuet}}, \bibinfo
  {author} {\bibfnamefont {H.}~\bibnamefont {Ruf}}, \bibinfo {author}
  {\bibfnamefont {A.}~\bibnamefont {Dubrouil}}, \bibinfo {author}
  {\bibfnamefont {S.}~\bibnamefont {Patchkovskii}}, \bibinfo {author}
  {\bibfnamefont {M.}~\bibnamefont {Spanner}}, \bibinfo {author} {\bibfnamefont
  {Y.}~\bibnamefont {Mairesse}}, \bibinfo {author} {\bibfnamefont
  {V.}~\bibnamefont {Blanchet}},  \emph {et~al.},\ }\href@noop {} {\bibfield
  {journal} {\bibinfo  {journal} {Science}\ }\textbf {\bibinfo {volume}
  {334}},\ \bibinfo {pages} {208} (\bibinfo {year} {2011})}\BibitemShut
  {NoStop}%
\bibitem [{\citenamefont {Niedziela}\ \emph {et~al.}(2019)\citenamefont
  {Niedziela}, \citenamefont {Bansal}, \citenamefont {May}, \citenamefont
  {Ding}, \citenamefont {T.L.-Atkins}, \citenamefont {Ehlers}, \citenamefont
  {Abernathy}, \citenamefont {Said},\ and\ \citenamefont
  {Delaire}}]{Niedziela_2019}%
  \BibitemOpen
  \bibfield  {author} {\bibinfo {author} {\bibfnamefont {J.}~\bibnamefont
  {Niedziela}}, \bibinfo {author} {\bibfnamefont {D.}~\bibnamefont {Bansal}},
  \bibinfo {author} {\bibfnamefont {A.}~\bibnamefont {May}}, \bibinfo {author}
  {\bibfnamefont {J.}~\bibnamefont {Ding}}, \bibinfo {author} {\bibnamefont
  {T.L.-Atkins}}, \bibinfo {author} {\bibfnamefont {G.}~\bibnamefont {Ehlers}},
  \bibinfo {author} {\bibfnamefont {D.}~\bibnamefont {Abernathy}}, \bibinfo
  {author} {\bibfnamefont {A.}~\bibnamefont {Said}}, \ and\ \bibinfo {author}
  {\bibfnamefont {O.}~\bibnamefont {Delaire}},\ }\href@noop {} {\bibfield
  {journal} {\bibinfo  {journal} {Nature Physics}\ }\textbf {\bibinfo {volume}
  {15}},\ \bibinfo {pages} {73} (\bibinfo {year} {2019})}\BibitemShut {NoStop}%
\bibitem [{\citenamefont {Dove}(1993)}]{Dove1993}%
  \BibitemOpen
  \bibfield  {author} {\bibinfo {author} {\bibfnamefont {M.}~\bibnamefont
  {Dove}},\ }\href@noop {} {\emph {\bibinfo {title} {Introduction to Lattice
  Dynamics}}}\ (\bibinfo  {publisher} {Cambridge University Press},\ \bibinfo
  {address} {Cambridge},\ \bibinfo {year} {1993})\BibitemShut {NoStop}%
\bibitem [{\citenamefont {Katsuki}\ \emph {et~al.}(2013)\citenamefont
  {Katsuki}, \citenamefont {Delagnes}, \citenamefont {Hosaka}, \citenamefont
  {Ishioka}, \citenamefont {Chiba}, \citenamefont {Zijlstra}, \citenamefont
  {Garcia}, \citenamefont {Takahashi}, \citenamefont {Watanabe}, \citenamefont
  {Kitajima} \emph {et~al.}}]{katsuki2013all}%
  \BibitemOpen
  \bibfield  {author} {\bibinfo {author} {\bibfnamefont {H.}~\bibnamefont
  {Katsuki}}, \bibinfo {author} {\bibfnamefont {J.}~\bibnamefont {Delagnes}},
  \bibinfo {author} {\bibfnamefont {K.}~\bibnamefont {Hosaka}}, \bibinfo
  {author} {\bibfnamefont {K.}~\bibnamefont {Ishioka}}, \bibinfo {author}
  {\bibfnamefont {H.}~\bibnamefont {Chiba}}, \bibinfo {author} {\bibfnamefont
  {E.}~\bibnamefont {Zijlstra}}, \bibinfo {author} {\bibfnamefont
  {M.}~\bibnamefont {Garcia}}, \bibinfo {author} {\bibfnamefont
  {H.}~\bibnamefont {Takahashi}}, \bibinfo {author} {\bibfnamefont
  {K.}~\bibnamefont {Watanabe}}, \bibinfo {author} {\bibfnamefont
  {M.}~\bibnamefont {Kitajima}},  \emph {et~al.},\ }\href@noop {} {\bibfield
  {journal} {\bibinfo  {journal} {Nature communications}\ }\textbf {\bibinfo
  {volume} {4}},\ \bibinfo {pages} {1} (\bibinfo {year} {2013})}\BibitemShut
  {NoStop}%
\bibitem [{\citenamefont {Bansal}\ \emph {et~al.}(2020)\citenamefont {Bansal},
  \citenamefont {Niedziela}, \citenamefont {Calder}, \citenamefont
  {Lanigan-Atkins}, \citenamefont {Rawl}, \citenamefont {Said}, \citenamefont
  {Abernathy}, \citenamefont {Kolesnikov}, \citenamefont {Zhou},\ and\
  \citenamefont {Delaire}}]{bansal2020magnetically}%
  \BibitemOpen
  \bibfield  {author} {\bibinfo {author} {\bibfnamefont {D.}~\bibnamefont
  {Bansal}}, \bibinfo {author} {\bibfnamefont {J.~L.}\ \bibnamefont
  {Niedziela}}, \bibinfo {author} {\bibfnamefont {S.}~\bibnamefont {Calder}},
  \bibinfo {author} {\bibfnamefont {T.}~\bibnamefont {Lanigan-Atkins}},
  \bibinfo {author} {\bibfnamefont {R.}~\bibnamefont {Rawl}}, \bibinfo {author}
  {\bibfnamefont {A.~H.}\ \bibnamefont {Said}}, \bibinfo {author}
  {\bibfnamefont {D.~L.}\ \bibnamefont {Abernathy}}, \bibinfo {author}
  {\bibfnamefont {A.~I.}\ \bibnamefont {Kolesnikov}}, \bibinfo {author}
  {\bibfnamefont {H.}~\bibnamefont {Zhou}}, \ and\ \bibinfo {author}
  {\bibfnamefont {O.}~\bibnamefont {Delaire}},\ }\href@noop {} {\bibfield
  {journal} {\bibinfo  {journal} {Nature Physics}\ }\textbf {\bibinfo {volume}
  {16}},\ \bibinfo {pages} {669} (\bibinfo {year} {2020})}\BibitemShut
  {NoStop}%
\bibitem [{\citenamefont {Hase}\ \emph {et~al.}(2015)\citenamefont {Hase},
  \citenamefont {Fons}, \citenamefont {Mitrofanov}, \citenamefont {Kolobov},\
  and\ \citenamefont {Tominaga}}]{hase2015femtosecond}%
  \BibitemOpen
  \bibfield  {author} {\bibinfo {author} {\bibfnamefont {M.}~\bibnamefont
  {Hase}}, \bibinfo {author} {\bibfnamefont {P.}~\bibnamefont {Fons}}, \bibinfo
  {author} {\bibfnamefont {K.}~\bibnamefont {Mitrofanov}}, \bibinfo {author}
  {\bibfnamefont {A.~V.}\ \bibnamefont {Kolobov}}, \ and\ \bibinfo {author}
  {\bibfnamefont {J.}~\bibnamefont {Tominaga}},\ }\href@noop {} {\bibfield
  {journal} {\bibinfo  {journal} {Nature communications}\ }\textbf {\bibinfo
  {volume} {6}},\ \bibinfo {pages} {1} (\bibinfo {year} {2015})}\BibitemShut
  {NoStop}%
\bibitem [{\citenamefont {Bansal}\ \emph {et~al.}(2018)\citenamefont {Bansal},
  \citenamefont {Niedziela}, \citenamefont {Sinclair}, \citenamefont {Garlea},
  \citenamefont {Abernathy}, \citenamefont {Chi}, \citenamefont {Ren},
  \citenamefont {Zhou},\ and\ \citenamefont {Delaire}}]{Bansal_2018}%
  \BibitemOpen
  \bibfield  {author} {\bibinfo {author} {\bibfnamefont {D.}~\bibnamefont
  {Bansal}}, \bibinfo {author} {\bibfnamefont {J.}~\bibnamefont {Niedziela}},
  \bibinfo {author} {\bibfnamefont {R.}~\bibnamefont {Sinclair}}, \bibinfo
  {author} {\bibfnamefont {V.}~\bibnamefont {Garlea}}, \bibinfo {author}
  {\bibfnamefont {D.}~\bibnamefont {Abernathy}}, \bibinfo {author}
  {\bibfnamefont {S.}~\bibnamefont {Chi}}, \bibinfo {author} {\bibfnamefont
  {Y.}~\bibnamefont {Ren}}, \bibinfo {author} {\bibfnamefont {H.}~\bibnamefont
  {Zhou}}, \ and\ \bibinfo {author} {\bibfnamefont {O.}~\bibnamefont
  {Delaire}},\ }\href@noop {} {\bibfield  {journal} {\bibinfo  {journal}
  {Nature Communications}\ }\textbf {\bibinfo {volume} {9}},\ \bibinfo {pages}
  {15} (\bibinfo {year} {2018})}\BibitemShut {NoStop}%
\bibitem [{\citenamefont {Fultz}(2010)}]{Fultz2010}%
  \BibitemOpen
  \bibfield  {author} {\bibinfo {author} {\bibfnamefont {B.}~\bibnamefont
  {Fultz}},\ }\href@noop {} {\bibfield  {journal} {\bibinfo  {journal}
  {Progress in Materials Science}\ }\textbf {\bibinfo {volume} {55}},\ \bibinfo
  {pages} {247} (\bibinfo {year} {2010})}\BibitemShut {NoStop}%
\bibitem [{\citenamefont {Gambetta}\ \emph {et~al.}(2006)\citenamefont
  {Gambetta}, \citenamefont {Manzoni}, \citenamefont {Menna}, \citenamefont
  {Meneghetti}, \citenamefont {Cerullo}, \citenamefont {Lanzani}, \citenamefont
  {Tretiak}, \citenamefont {Piryatinski}, \citenamefont {Saxena}, \citenamefont
  {Martin} \emph {et~al.}}]{gambetta2006real}%
  \BibitemOpen
  \bibfield  {author} {\bibinfo {author} {\bibfnamefont {A.}~\bibnamefont
  {Gambetta}}, \bibinfo {author} {\bibfnamefont {C.}~\bibnamefont {Manzoni}},
  \bibinfo {author} {\bibfnamefont {E.}~\bibnamefont {Menna}}, \bibinfo
  {author} {\bibfnamefont {M.}~\bibnamefont {Meneghetti}}, \bibinfo {author}
  {\bibfnamefont {G.}~\bibnamefont {Cerullo}}, \bibinfo {author} {\bibfnamefont
  {G.}~\bibnamefont {Lanzani}}, \bibinfo {author} {\bibfnamefont
  {S.}~\bibnamefont {Tretiak}}, \bibinfo {author} {\bibfnamefont
  {A.}~\bibnamefont {Piryatinski}}, \bibinfo {author} {\bibfnamefont
  {A.}~\bibnamefont {Saxena}}, \bibinfo {author} {\bibfnamefont
  {R.}~\bibnamefont {Martin}},  \emph {et~al.},\ }\href@noop {} {\bibfield
  {journal} {\bibinfo  {journal} {Nature Physics}\ }\textbf {\bibinfo {volume}
  {2}},\ \bibinfo {pages} {515} (\bibinfo {year} {2006})}\BibitemShut {NoStop}%
\bibitem [{\citenamefont {Dhar}\ \emph {et~al.}(1994)\citenamefont {Dhar},
  \citenamefont {Rogers},\ and\ \citenamefont {Nelson}}]{dhar1994time}%
  \BibitemOpen
  \bibfield  {author} {\bibinfo {author} {\bibfnamefont {L.}~\bibnamefont
  {Dhar}}, \bibinfo {author} {\bibfnamefont {J.~A.}\ \bibnamefont {Rogers}}, \
  and\ \bibinfo {author} {\bibfnamefont {K.~A.}\ \bibnamefont {Nelson}},\
  }\href@noop {} {\bibfield  {journal} {\bibinfo  {journal} {Chemical Reviews}\
  }\textbf {\bibinfo {volume} {94}},\ \bibinfo {pages} {157} (\bibinfo {year}
  {1994})}\BibitemShut {NoStop}%
\bibitem [{\citenamefont {Debnath}\ \emph {et~al.}(2021)\citenamefont
  {Debnath}, \citenamefont {Sarker}, \citenamefont {Huang}, \citenamefont
  {Han}, \citenamefont {Dey}, \citenamefont {Polavarapu}, \citenamefont
  {Levchenko},\ and\ \citenamefont {Feldmann}}]{debnath2021coherent}%
  \BibitemOpen
  \bibfield  {author} {\bibinfo {author} {\bibfnamefont {T.}~\bibnamefont
  {Debnath}}, \bibinfo {author} {\bibfnamefont {D.}~\bibnamefont {Sarker}},
  \bibinfo {author} {\bibfnamefont {H.}~\bibnamefont {Huang}}, \bibinfo
  {author} {\bibfnamefont {Z.-K.}\ \bibnamefont {Han}}, \bibinfo {author}
  {\bibfnamefont {A.}~\bibnamefont {Dey}}, \bibinfo {author} {\bibfnamefont
  {L.}~\bibnamefont {Polavarapu}}, \bibinfo {author} {\bibfnamefont {S.~V.}\
  \bibnamefont {Levchenko}}, \ and\ \bibinfo {author} {\bibfnamefont
  {J.}~\bibnamefont {Feldmann}},\ }\href@noop {} {\bibfield  {journal}
  {\bibinfo  {journal} {Nature communications}\ }\textbf {\bibinfo {volume}
  {12}},\ \bibinfo {pages} {1} (\bibinfo {year} {2021})}\BibitemShut {NoStop}%
\bibitem [{\citenamefont {Graf}\ \emph {et~al.}(2007)\citenamefont {Graf},
  \citenamefont {Molitor}, \citenamefont {Ensslin}, \citenamefont {Stampfer},
  \citenamefont {Jungen}, \citenamefont {Hierold},\ and\ \citenamefont
  {Wirtz}}]{graf2007spatially}%
  \BibitemOpen
  \bibfield  {author} {\bibinfo {author} {\bibfnamefont {D.}~\bibnamefont
  {Graf}}, \bibinfo {author} {\bibfnamefont {F.}~\bibnamefont {Molitor}},
  \bibinfo {author} {\bibfnamefont {K.}~\bibnamefont {Ensslin}}, \bibinfo
  {author} {\bibfnamefont {C.}~\bibnamefont {Stampfer}}, \bibinfo {author}
  {\bibfnamefont {A.}~\bibnamefont {Jungen}}, \bibinfo {author} {\bibfnamefont
  {C.}~\bibnamefont {Hierold}}, \ and\ \bibinfo {author} {\bibfnamefont
  {L.}~\bibnamefont {Wirtz}},\ }\href@noop {} {\bibfield  {journal} {\bibinfo
  {journal} {Nano Letters}\ }\textbf {\bibinfo {volume} {7}},\ \bibinfo {pages}
  {238} (\bibinfo {year} {2007})}\BibitemShut {NoStop}%
\bibitem [{\citenamefont {Virga}\ \emph {et~al.}(2019)\citenamefont {Virga},
  \citenamefont {Ferrante}, \citenamefont {Batignani}, \citenamefont
  {De~Fazio}, \citenamefont {Nunn}, \citenamefont {Ferrari}, \citenamefont
  {Cerullo},\ and\ \citenamefont {Scopigno}}]{virga2019coherent}%
  \BibitemOpen
  \bibfield  {author} {\bibinfo {author} {\bibfnamefont {A.}~\bibnamefont
  {Virga}}, \bibinfo {author} {\bibfnamefont {C.}~\bibnamefont {Ferrante}},
  \bibinfo {author} {\bibfnamefont {G.}~\bibnamefont {Batignani}}, \bibinfo
  {author} {\bibfnamefont {D.}~\bibnamefont {De~Fazio}}, \bibinfo {author}
  {\bibfnamefont {A.}~\bibnamefont {Nunn}}, \bibinfo {author} {\bibfnamefont
  {A.}~\bibnamefont {Ferrari}}, \bibinfo {author} {\bibfnamefont
  {G.}~\bibnamefont {Cerullo}}, \ and\ \bibinfo {author} {\bibfnamefont
  {T.}~\bibnamefont {Scopigno}},\ }\href@noop {} {\bibfield  {journal}
  {\bibinfo  {journal} {Nature Communications}\ }\textbf {\bibinfo {volume}
  {10}},\ \bibinfo {pages} {1} (\bibinfo {year} {2019})}\BibitemShut {NoStop}%
\bibitem [{\citenamefont {Koivistoinen}\ \emph {et~al.}(2017)\citenamefont
  {Koivistoinen}, \citenamefont {Myllyperkio},\ and\ \citenamefont
  {Pettersson}}]{koivistoinen2017time}%
  \BibitemOpen
  \bibfield  {author} {\bibinfo {author} {\bibfnamefont {J.}~\bibnamefont
  {Koivistoinen}}, \bibinfo {author} {\bibfnamefont {P.}~\bibnamefont
  {Myllyperkio}}, \ and\ \bibinfo {author} {\bibfnamefont {M.}~\bibnamefont
  {Pettersson}},\ }\href@noop {} {\bibfield  {journal} {\bibinfo  {journal}
  {Journal of Physical Chemistry Letters}\ }\textbf {\bibinfo {volume} {8}},\
  \bibinfo {pages} {4108} (\bibinfo {year} {2017})}\BibitemShut {NoStop}%
\bibitem [{\citenamefont {Rana}\ \emph {et~al.}(2021)\citenamefont {Rana},
  \citenamefont {Roy}, \citenamefont {Bansal},\ and\ \citenamefont
  {Dixit}}]{rana2021four}%
  \BibitemOpen
  \bibfield  {author} {\bibinfo {author} {\bibfnamefont {N.}~\bibnamefont
  {Rana}}, \bibinfo {author} {\bibfnamefont {A.~P.}\ \bibnamefont {Roy}},
  \bibinfo {author} {\bibfnamefont {D.}~\bibnamefont {Bansal}}, \ and\ \bibinfo
  {author} {\bibfnamefont {G.}~\bibnamefont {Dixit}},\ }\href@noop {}
  {\bibfield  {journal} {\bibinfo  {journal} {npj Computational Materials}\
  }\textbf {\bibinfo {volume} {7}},\ \bibinfo {pages} {1} (\bibinfo {year}
  {2021})}\BibitemShut {NoStop}%
\bibitem [{\citenamefont {Brown}\ \emph {et~al.}(2019)\citenamefont {Brown},
  \citenamefont {Gleason}, \citenamefont {Galtier}, \citenamefont
  {Higginbotham}, \citenamefont {Arnold}, \citenamefont {Fry}, \citenamefont
  {Granados}, \citenamefont {Hashim}, \citenamefont {Schroer}, \citenamefont
  {Schropp} \emph {et~al.}}]{brown2019direct}%
  \BibitemOpen
  \bibfield  {author} {\bibinfo {author} {\bibfnamefont {S.~B.}\ \bibnamefont
  {Brown}}, \bibinfo {author} {\bibfnamefont {A.}~\bibnamefont {Gleason}},
  \bibinfo {author} {\bibfnamefont {E.}~\bibnamefont {Galtier}}, \bibinfo
  {author} {\bibfnamefont {A.}~\bibnamefont {Higginbotham}}, \bibinfo {author}
  {\bibfnamefont {B.}~\bibnamefont {Arnold}}, \bibinfo {author} {\bibfnamefont
  {A.}~\bibnamefont {Fry}}, \bibinfo {author} {\bibfnamefont {E.}~\bibnamefont
  {Granados}}, \bibinfo {author} {\bibfnamefont {A.}~\bibnamefont {Hashim}},
  \bibinfo {author} {\bibfnamefont {C.~G.}\ \bibnamefont {Schroer}}, \bibinfo
  {author} {\bibfnamefont {A.}~\bibnamefont {Schropp}},  \emph {et~al.},\
  }\href@noop {} {\bibfield  {journal} {\bibinfo  {journal} {Science advances}\
  }\textbf {\bibinfo {volume} {5}},\ \bibinfo {pages} {eaau8044} (\bibinfo
  {year} {2019})}\BibitemShut {NoStop}%
\bibitem [{\citenamefont {Flannigan}(2018)}]{flannigan2018electrons}%
  \BibitemOpen
  \bibfield  {author} {\bibinfo {author} {\bibfnamefont {D.~J.}\ \bibnamefont
  {Flannigan}},\ }\href@noop {} {\bibfield  {journal} {\bibinfo  {journal}
  {Physics}\ }\textbf {\bibinfo {volume} {11}},\ \bibinfo {pages} {53}
  (\bibinfo {year} {2018})}\BibitemShut {NoStop}%
\bibitem [{\citenamefont {Gierz}\ \emph {et~al.}(2015)\citenamefont {Gierz},
  \citenamefont {Mitrano}, \citenamefont {Bromberger}, \citenamefont {Cacho},
  \citenamefont {Chapman}, \citenamefont {Springate}, \citenamefont {Link},
  \citenamefont {Starke}, \citenamefont {Sachs}, \citenamefont {Eckstein} \emph
  {et~al.}}]{gierz2015phonon}%
  \BibitemOpen
  \bibfield  {author} {\bibinfo {author} {\bibfnamefont {I.}~\bibnamefont
  {Gierz}}, \bibinfo {author} {\bibfnamefont {M.}~\bibnamefont {Mitrano}},
  \bibinfo {author} {\bibfnamefont {H.}~\bibnamefont {Bromberger}}, \bibinfo
  {author} {\bibfnamefont {C.}~\bibnamefont {Cacho}}, \bibinfo {author}
  {\bibfnamefont {R.}~\bibnamefont {Chapman}}, \bibinfo {author} {\bibfnamefont
  {E.}~\bibnamefont {Springate}}, \bibinfo {author} {\bibfnamefont
  {S.}~\bibnamefont {Link}}, \bibinfo {author} {\bibfnamefont {U.}~\bibnamefont
  {Starke}}, \bibinfo {author} {\bibfnamefont {B.}~\bibnamefont {Sachs}},
  \bibinfo {author} {\bibfnamefont {M.}~\bibnamefont {Eckstein}},  \emph
  {et~al.},\ }\href@noop {} {\bibfield  {journal} {\bibinfo  {journal}
  {Physical Review Letters}\ }\textbf {\bibinfo {volume} {114}},\ \bibinfo
  {pages} {125503} (\bibinfo {year} {2015})}\BibitemShut {NoStop}%
\bibitem [{\citenamefont {Moulet}\ \emph {et~al.}(2017)\citenamefont {Moulet},
  \citenamefont {Bertrand}, \citenamefont {Klostermann}, \citenamefont
  {Guggenmos}, \citenamefont {Karpowicz},\ and\ \citenamefont
  {Goulielmakis}}]{moulet2017soft}%
  \BibitemOpen
  \bibfield  {author} {\bibinfo {author} {\bibfnamefont {A.}~\bibnamefont
  {Moulet}}, \bibinfo {author} {\bibfnamefont {J.~B.}\ \bibnamefont
  {Bertrand}}, \bibinfo {author} {\bibfnamefont {T.}~\bibnamefont
  {Klostermann}}, \bibinfo {author} {\bibfnamefont {A.}~\bibnamefont
  {Guggenmos}}, \bibinfo {author} {\bibfnamefont {N.}~\bibnamefont
  {Karpowicz}}, \ and\ \bibinfo {author} {\bibfnamefont {E.}~\bibnamefont
  {Goulielmakis}},\ }\href@noop {} {\bibfield  {journal} {\bibinfo  {journal}
  {Science}\ }\textbf {\bibinfo {volume} {357}},\ \bibinfo {pages} {1134}
  (\bibinfo {year} {2017})}\BibitemShut {NoStop}%
\bibitem [{\citenamefont {G{\'e}neaux}\ \emph {et~al.}(2020)\citenamefont
  {G{\'e}neaux}, \citenamefont {Kaplan}, \citenamefont {Yue}, \citenamefont
  {Ross}, \citenamefont {B{\ae}kh{\o}j}, \citenamefont {Kraus}, \citenamefont
  {Chang}, \citenamefont {Guggenmos}, \citenamefont {Huang}, \citenamefont
  {Z{\"u}rch} \emph {et~al.}}]{geneaux2020attosecond}%
  \BibitemOpen
  \bibfield  {author} {\bibinfo {author} {\bibfnamefont {R.}~\bibnamefont
  {G{\'e}neaux}}, \bibinfo {author} {\bibfnamefont {C.~J.}\ \bibnamefont
  {Kaplan}}, \bibinfo {author} {\bibfnamefont {L.}~\bibnamefont {Yue}},
  \bibinfo {author} {\bibfnamefont {A.~D.}\ \bibnamefont {Ross}}, \bibinfo
  {author} {\bibfnamefont {J.~E.}\ \bibnamefont {B{\ae}kh{\o}j}}, \bibinfo
  {author} {\bibfnamefont {P.~M.}\ \bibnamefont {Kraus}}, \bibinfo {author}
  {\bibfnamefont {H.-T.}\ \bibnamefont {Chang}}, \bibinfo {author}
  {\bibfnamefont {A.}~\bibnamefont {Guggenmos}}, \bibinfo {author}
  {\bibfnamefont {M.-Y.}\ \bibnamefont {Huang}}, \bibinfo {author}
  {\bibfnamefont {M.}~\bibnamefont {Z{\"u}rch}},  \emph {et~al.},\ }\href@noop
  {} {\bibfield  {journal} {\bibinfo  {journal} {Physical Review Letters}\
  }\textbf {\bibinfo {volume} {124}},\ \bibinfo {pages} {207401} (\bibinfo
  {year} {2020})}\BibitemShut {NoStop}%
\bibitem [{\citenamefont {Kim}\ \emph {et~al.}(2013)\citenamefont {Kim},
  \citenamefont {Nugraha}, \citenamefont {Booshehri}, \citenamefont
  {H{\'a}roz}, \citenamefont {Sato}, \citenamefont {Sanders}, \citenamefont
  {Yee}, \citenamefont {Lim}, \citenamefont {Stanton}, \citenamefont {Saito}
  \emph {et~al.}}]{kim2013coherent}%
  \BibitemOpen
  \bibfield  {author} {\bibinfo {author} {\bibfnamefont {J.-H.}\ \bibnamefont
  {Kim}}, \bibinfo {author} {\bibfnamefont {A.}~\bibnamefont {Nugraha}},
  \bibinfo {author} {\bibfnamefont {L.}~\bibnamefont {Booshehri}}, \bibinfo
  {author} {\bibfnamefont {E.}~\bibnamefont {H{\'a}roz}}, \bibinfo {author}
  {\bibfnamefont {K.}~\bibnamefont {Sato}}, \bibinfo {author} {\bibfnamefont
  {G.}~\bibnamefont {Sanders}}, \bibinfo {author} {\bibfnamefont {K.-J.}\
  \bibnamefont {Yee}}, \bibinfo {author} {\bibfnamefont {Y.-S.}\ \bibnamefont
  {Lim}}, \bibinfo {author} {\bibfnamefont {C.}~\bibnamefont {Stanton}},
  \bibinfo {author} {\bibfnamefont {R.}~\bibnamefont {Saito}},  \emph
  {et~al.},\ }\href@noop {} {\bibfield  {journal} {\bibinfo  {journal}
  {Chemical Physics}\ }\textbf {\bibinfo {volume} {413}},\ \bibinfo {pages}
  {55} (\bibinfo {year} {2013})}\BibitemShut {NoStop}%
\bibitem [{\citenamefont {Mohanty}\ and\ \citenamefont
  {Heller}(2019)}]{mohanty2019lazy}%
  \BibitemOpen
  \bibfield  {author} {\bibinfo {author} {\bibfnamefont {V.}~\bibnamefont
  {Mohanty}}\ and\ \bibinfo {author} {\bibfnamefont {E.~J.}\ \bibnamefont
  {Heller}},\ }\href@noop {} {\bibfield  {journal} {\bibinfo  {journal}
  {Proceedings of the National Academy of Sciences}\ }\textbf {\bibinfo
  {volume} {116}},\ \bibinfo {pages} {18316} (\bibinfo {year}
  {2019})}\BibitemShut {NoStop}%
\bibitem [{\citenamefont {Rodriguez-Vega}\ \emph {et~al.}(2021)\citenamefont
  {Rodriguez-Vega}, \citenamefont {Vogl},\ and\ \citenamefont
  {Fiete}}]{rodriguez2021direct}%
  \BibitemOpen
  \bibfield  {author} {\bibinfo {author} {\bibfnamefont {M.}~\bibnamefont
  {Rodriguez-Vega}}, \bibinfo {author} {\bibfnamefont {M.}~\bibnamefont
  {Vogl}}, \ and\ \bibinfo {author} {\bibfnamefont {G.~A.}\ \bibnamefont
  {Fiete}},\ }\href@noop {} {\bibfield  {journal} {\bibinfo  {journal}
  {Physical Review B}\ }\textbf {\bibinfo {volume} {104}},\ \bibinfo {pages}
  {245135} (\bibinfo {year} {2021})}\BibitemShut {NoStop}%
\bibitem [{\citenamefont {Wang}\ and\ \citenamefont
  {Fischer}(2014)}]{wang2014topological}%
  \BibitemOpen
  \bibfield  {author} {\bibinfo {author} {\bibfnamefont {J.}~\bibnamefont
  {Wang}}\ and\ \bibinfo {author} {\bibfnamefont {S.}~\bibnamefont {Fischer}},\
  }\href@noop {} {\bibfield  {journal} {\bibinfo  {journal} {Physical Review
  B}\ }\textbf {\bibinfo {volume} {89}},\ \bibinfo {pages} {245421} (\bibinfo
  {year} {2014})}\BibitemShut {NoStop}%
\bibitem [{\citenamefont {Moon}\ and\ \citenamefont
  {Koshino}(2013)}]{moon2013optical}%
  \BibitemOpen
  \bibfield  {author} {\bibinfo {author} {\bibfnamefont {P.}~\bibnamefont
  {Moon}}\ and\ \bibinfo {author} {\bibfnamefont {M.}~\bibnamefont {Koshino}},\
  }\href@noop {} {\bibfield  {journal} {\bibinfo  {journal} {Physical Review
  B}\ }\textbf {\bibinfo {volume} {87}},\ \bibinfo {pages} {205404} (\bibinfo
  {year} {2013})}\BibitemShut {NoStop}%
\bibitem [{\citenamefont {Currie}\ \emph {et~al.}(2011)\citenamefont {Currie},
  \citenamefont {Caldwell}, \citenamefont {Bezares}, \citenamefont {Robinson},
  \citenamefont {Anderson}, \citenamefont {Chun},\ and\ \citenamefont
  {Tadjer}}]{currie2011quantifying}%
  \BibitemOpen
  \bibfield  {author} {\bibinfo {author} {\bibfnamefont {M.}~\bibnamefont
  {Currie}}, \bibinfo {author} {\bibfnamefont {J.~D.}\ \bibnamefont
  {Caldwell}}, \bibinfo {author} {\bibfnamefont {F.~J.}\ \bibnamefont
  {Bezares}}, \bibinfo {author} {\bibfnamefont {J.}~\bibnamefont {Robinson}},
  \bibinfo {author} {\bibfnamefont {T.}~\bibnamefont {Anderson}}, \bibinfo
  {author} {\bibfnamefont {H.}~\bibnamefont {Chun}}, \ and\ \bibinfo {author}
  {\bibfnamefont {M.}~\bibnamefont {Tadjer}},\ }\href@noop {} {\bibfield
  {journal} {\bibinfo  {journal} {Applied Physics Letters}\ }\textbf {\bibinfo
  {volume} {99}},\ \bibinfo {pages} {211909} (\bibinfo {year}
  {2011})}\BibitemShut {NoStop}%
\bibitem [{\citenamefont {Heide}\ \emph {et~al.}(2018)\citenamefont {Heide},
  \citenamefont {Higuchi}, \citenamefont {Weber},\ and\ \citenamefont
  {Hommelhoff}}]{heide2018coherent}%
  \BibitemOpen
  \bibfield  {author} {\bibinfo {author} {\bibfnamefont {C.}~\bibnamefont
  {Heide}}, \bibinfo {author} {\bibfnamefont {T.}~\bibnamefont {Higuchi}},
  \bibinfo {author} {\bibfnamefont {H.~B.}\ \bibnamefont {Weber}}, \ and\
  \bibinfo {author} {\bibfnamefont {P.}~\bibnamefont {Hommelhoff}},\
  }\href@noop {} {\bibfield  {journal} {\bibinfo  {journal} {Physical Review
  Letters}\ }\textbf {\bibinfo {volume} {121}},\ \bibinfo {pages} {207401}
  (\bibinfo {year} {2018})}\BibitemShut {NoStop}%
\bibitem [{\citenamefont {Higuchi}\ \emph {et~al.}(2017)\citenamefont
  {Higuchi}, \citenamefont {Heide}, \citenamefont {Ullmann}, \citenamefont
  {Weber},\ and\ \citenamefont {Hommelhoff}}]{higuchi2017light}%
  \BibitemOpen
  \bibfield  {author} {\bibinfo {author} {\bibfnamefont {T.}~\bibnamefont
  {Higuchi}}, \bibinfo {author} {\bibfnamefont {C.}~\bibnamefont {Heide}},
  \bibinfo {author} {\bibfnamefont {K.}~\bibnamefont {Ullmann}}, \bibinfo
  {author} {\bibfnamefont {H.~B.}\ \bibnamefont {Weber}}, \ and\ \bibinfo
  {author} {\bibfnamefont {P.}~\bibnamefont {Hommelhoff}},\ }\href@noop {}
  {\bibfield  {journal} {\bibinfo  {journal} {Nature}\ }\textbf {\bibinfo
  {volume} {550}},\ \bibinfo {pages} {224} (\bibinfo {year}
  {2017})}\BibitemShut {NoStop}%
\bibitem [{\citenamefont {Yoshikawa}\ \emph {et~al.}(2017)\citenamefont
  {Yoshikawa}, \citenamefont {Tamaya},\ and\ \citenamefont
  {Tanaka}}]{yoshikawa2017high}%
  \BibitemOpen
  \bibfield  {author} {\bibinfo {author} {\bibfnamefont {N.}~\bibnamefont
  {Yoshikawa}}, \bibinfo {author} {\bibfnamefont {T.}~\bibnamefont {Tamaya}}, \
  and\ \bibinfo {author} {\bibfnamefont {K.}~\bibnamefont {Tanaka}},\
  }\href@noop {} {\bibfield  {journal} {\bibinfo  {journal} {Science}\ }\textbf
  {\bibinfo {volume} {356}},\ \bibinfo {pages} {736} (\bibinfo {year}
  {2017})}\BibitemShut {NoStop}%
\bibitem [{\citenamefont {Mrudul}\ and\ \citenamefont
  {Dixit}(2021{\natexlab{b}})}]{mrudul2021high}%
  \BibitemOpen
  \bibfield  {author} {\bibinfo {author} {\bibfnamefont {M.}~\bibnamefont
  {Mrudul}}\ and\ \bibinfo {author} {\bibfnamefont {G.}~\bibnamefont {Dixit}},\
  }\href@noop {} {\bibfield  {journal} {\bibinfo  {journal} {Physical Review
  B}\ }\textbf {\bibinfo {volume} {103}},\ \bibinfo {pages} {094308} (\bibinfo
  {year} {2021}{\natexlab{b}})}\BibitemShut {NoStop}%
\bibitem [{\citenamefont {Al-Naib}\ \emph {et~al.}(2014)\citenamefont
  {Al-Naib}, \citenamefont {Sipe},\ and\ \citenamefont {Dignam}}]{al2014high}%
  \BibitemOpen
  \bibfield  {author} {\bibinfo {author} {\bibfnamefont {I.}~\bibnamefont
  {Al-Naib}}, \bibinfo {author} {\bibfnamefont {J.~E.}\ \bibnamefont {Sipe}}, \
  and\ \bibinfo {author} {\bibfnamefont {M.~M.}\ \bibnamefont {Dignam}},\
  }\href@noop {} {\bibfield  {journal} {\bibinfo  {journal} {Physical Review
  B}\ }\textbf {\bibinfo {volume} {90}},\ \bibinfo {pages} {245423} (\bibinfo
  {year} {2014})}\BibitemShut {NoStop}%
\bibitem [{\citenamefont {Du}\ \emph {et~al.}(2021)\citenamefont {Du},
  \citenamefont {Liu}, \citenamefont {Zeng},\ and\ \citenamefont
  {Li}}]{du2021high}%
  \BibitemOpen
  \bibfield  {author} {\bibinfo {author} {\bibfnamefont {M.}~\bibnamefont
  {Du}}, \bibinfo {author} {\bibfnamefont {C.}~\bibnamefont {Liu}}, \bibinfo
  {author} {\bibfnamefont {Z.}~\bibnamefont {Zeng}}, \ and\ \bibinfo {author}
  {\bibfnamefont {R.}~\bibnamefont {Li}},\ }\href@noop {} {\bibfield  {journal}
  {\bibinfo  {journal} {Physical Review A}\ }\textbf {\bibinfo {volume}
  {104}},\ \bibinfo {pages} {033113} (\bibinfo {year} {2021})}\BibitemShut
  {NoStop}%
\bibitem [{\citenamefont {Neufeld}\ \emph {et~al.}(2019)\citenamefont
  {Neufeld}, \citenamefont {Podolsky},\ and\ \citenamefont
  {Cohen}}]{neufeld2019floquet}%
  \BibitemOpen
  \bibfield  {author} {\bibinfo {author} {\bibfnamefont {O.}~\bibnamefont
  {Neufeld}}, \bibinfo {author} {\bibfnamefont {D.}~\bibnamefont {Podolsky}}, \
  and\ \bibinfo {author} {\bibfnamefont {O.}~\bibnamefont {Cohen}},\
  }\href@noop {} {\bibfield  {journal} {\bibinfo  {journal} {Nature
  communications}\ }\textbf {\bibinfo {volume} {10}},\ \bibinfo {pages} {1}
  (\bibinfo {year} {2019})}\BibitemShut {NoStop}%
\bibitem [{\citenamefont {Nagai}\ \emph {et~al.}(2020)\citenamefont {Nagai},
  \citenamefont {Uchida}, \citenamefont {Yoshikawa}, \citenamefont {Endo},
  \citenamefont {Miyata},\ and\ \citenamefont {Tanaka}}]{nagai2020dynamical}%
  \BibitemOpen
  \bibfield  {author} {\bibinfo {author} {\bibfnamefont {K.}~\bibnamefont
  {Nagai}}, \bibinfo {author} {\bibfnamefont {K.}~\bibnamefont {Uchida}},
  \bibinfo {author} {\bibfnamefont {N.}~\bibnamefont {Yoshikawa}}, \bibinfo
  {author} {\bibfnamefont {T.}~\bibnamefont {Endo}}, \bibinfo {author}
  {\bibfnamefont {Y.}~\bibnamefont {Miyata}}, \ and\ \bibinfo {author}
  {\bibfnamefont {K.}~\bibnamefont {Tanaka}},\ }\href@noop {} {\bibfield
  {journal} {\bibinfo  {journal} {Communications Physics}\ }\textbf {\bibinfo
  {volume} {3}},\ \bibinfo {pages} {1} (\bibinfo {year} {2020})}\BibitemShut
  {NoStop}%
\end{thebibliography}


\end{document}